\documentclass[12pt]{article}
\usepackage{graphicx,subfig,rotate}
\usepackage{amsmath,amssymb,latexsym}
\newcommand{\email}{E-mail address: }
\def \be{\begin{equation}}
\def \ee{\end{equation}}
\def \bea{\begin{eqnarray}}
\def \eea{\end{eqnarray}}

\def \Box{\text{box}}
\def \Count{\text{count}}
\def \Peng{\text{peng}}
\def \chargino{\tilde{\chi}^{\pm}}
\def \sneutrino{\tilde{\n}}

\def \upsquark{\tilde{u}}
\def \gev{{\text{GeV}}}
\def \nnu{\nonumber}
\def \diag{{\mathrm{diag}}}
\def \b{\beta}
\def \n{\nu}
\def \G{\Gamma}
\def \m{\mu}
\def \la{\lambda}

\def \bs{{B}_s \to \mu^+ \mu^-}
\def \bsm{b \to s \mu^+ \mu^-}
\def \br{Br(B_s \to \mu^+ \mu^-)}
\def \bl{\bar{B}_s \to l^+ l^-}

\def \tb{\tan\beta}
\def\lesssim{\mathrel{\hbox{\rlap{\hbox{\lower4pt\hbox{$\sim$}}}\hbox{$<$}}}} 
\def\gtrsim{\mathrel{\hbox{\rlap{\hbox{\lower4pt\hbox{$\sim$}}}\hbox{$>$}}}} 
\begin{document}
\begin{flushright}
TIFR/TH/09-11
\end{flushright}
\vskip 5pt
\begin{center}
{\Large{\bf $B_s \to \mu^+ \mu^-$ decay in the R-parity violating minimal supergravity}}
\vskip 25pt 
{\bf Ashutosh Kumar Alok$^\star$\footnote{\tt\email alok@theory.tifr.res.in}} and
{\bf Sudhir Kumar Gupta$^\dagger$\footnote{\tt\email skgupta@iastate.edu}} \\\vskip 10pt
{\em $^\star$\hspace{-.2cm}
Tata Institute of Fundamental Research\\ 
Homi Bhabha Road, Mumbai | 400005, INDIA\\[.2cm]
$^\dagger$\hspace{-.2cm}
Department of  Physics \&  Astronomy
\\Iowa  State University, Ames,  IA | 50011, USA}\\[.2cm]
\normalsize
\end{center}
\vskip 25pt
\begin{abstract}
 We study $B_s \to \mu^+ \mu^-$ in the context of the R-parity violating minimal
supergravity in the high $\tan\beta$ regime. We find that the lowest value of the branching ratio
can go well below the present LHCb sensitivity and hence $B_s \to \mu^+ \mu^-$ can even be invisible to the LHC. 
We also find that the present
upper bound on $Br(B_s \to \mu^+ \mu^-)$ puts strong constraint on the minimal supergravity
parameter space. The constraints become more severe if the upper bound
is close to its standard model prediction.
\end{abstract}

\newpage

\section{Introduction}

Flavor changing neutral current (FCNC) interactions serve as an important probe to test the standard model (SM) and its possible interactions. The reason is that these interactions are forbidden at the tree level within the SM and can  occur only via one or more loops. Hence such processes are highly suppressed within the SM. The high statistics experiments at the Large Hadron Collier (LHC) and the Super-B factories will measure FCNC interactions with high accuracy and hence will provide tests of higher order corrections to the SM and its possible extensions.

The quark level transition $\bsm$ serves as an important probe to test the SM at the loop level and also constrain many of its possible extensions. The $\bsm$ transition is responsible for the (i) inclusive semi-leptonic decay $B \to X_s \mu^+ \mu^-$, 
(ii) exclusive semi-leptonic decays $B \to
(K, K^*) \mu^+ \mu^-$, and (iii) purely leptonic decay $\bs$. Both the exclusive and inclusive semi-leptonic decays have been observed in experiments \cite{babar04_incl,belle05_incl,babar-03,babar-06,belle-03,hfag} with a branching ratio close to their SM predictions \cite{ali-02,lunghi,kruger-01,isidori,Asatryan:2001zw,Asatrian:2001de}.

The decay $\bs$ is helicity suppressed in the SM and  its branching ratio is predicted to be 
$(3.35\pm 0.32)\times 10^{-9}$ \cite{blanke,buchalla,buras-01}. This decay is yet to be observed experimentally. Recently the upper
bound on its branching ratio has been improved to \cite{cdf-07}
\begin{equation}
Br(\bs)<5.8 \times 10^{-8} \qquad (2\sigma\, \rm C.L.)
\end{equation}
which is still larger than an order of magnitude above its SM prediction. $\bs$ will be one of the important rare B 
decays to be studied at the LHC.
The LHCb can exclude the region between $10^{-8}$ and the SM prediction with very little luminosity $\sim 0.5\, fb^{-1}$. It has the potential for a 
$3 \sigma$ ($5 \sigma$) observation (discovery) of the SM prediction with $\sim 2\, fb^{-1}$ ($\sim 6\, fb^{-1}$) of data \cite{Lenzi:2007nq}. Hence with
a sensitivity exceeding the SM prediction of $\br$, LHCb will be able to observe both ehnancements as well as suppression in the branching ratio of  $\bs$.
The general purpose detectors, ATLAS and CMS can reconstruct the $\bs$ signal with $3 \sigma$ significance after three years of running at a luminosity of $10^{33}\,{\rm cm}^{-2}{\rm s}^{-1}$ \cite{Smizanska:2008qm}.

As $\bs$ is highly suppressed within the SM, it can serve as an 
important probe to test many new physics models. New physics in the form 
of magnetic dipole and tensor operators do not contribute to $\br$. In 
\cite{Alok:2005ep}, it was shown that new physics in the form of 
vector/axial-vector operators are constrained by the present measurement 
of the branching ratios of the exclusive semi-leptonic decays $B \to 
(K,K^*)\mu^+ \mu^-$ and an order of magnitude enhancement in $\br$ due 
to such operators is not possible. However if new physics is in the form 
of scalar/pseudoscalar operators then the measurement of $B \to K^* 
\mu^+ \mu^-$ fails to put any useful constraint on the new physics 
couplings and allows an order of magnitude or more enhancement in $\br$. 
Therefore $\br$ is sensitive to the new physics models with non standard 
scalar particles like the multi Higgs doublet models and supersymmetric 
models. This is the reason why the decay $\bs$ has been studied in 
literature in great detail in context of multi-Higgs doublet models as 
well as supersymmetry (SUSY) \cite{Skiba:1992mg}-\cite{Allanach:2009ne}.
SUSY  is among the leading candidates for the extensions of the SM. This besides providing a natural 
solution to electroweak hierarchy problem counts upon many interesting features such as suppression of flavor changing neutral current, provides a 
candidate for cold dark matter, common ingredient to superstrings/M-theory etc. Because of the broken nature of SUSY, it is yet to be observed in experiments.

In this paper we study $\bs$ in the context of R-parity violating (RPV) minimal supergravity (mSugra) framework for large $\tb$.  Though $\bs$ has been studied in RPV supersymmetry in \cite{Chen:2005kt,Xu:2006vk,Wang:2007sp}, these studies focus  only  on the contributions due to RPV terms only. Here, we present the contribution due to  the two Higgs doublet and R-parity conserving (RPC) terms along with RPV terms and study the enhancement as well as suppression of $\br$ as compared to its SM prediction. We also study the constrains on the mSugra parameter space coming from the present upper bound on the branching ratio of $\bs$. In addition, we also see how the constraints change if the upper bound on $\br$ is brought down to a value close to its SM prediction.

The paper is organized as follows. In Sec. \ref{bsbr}, we discuss $\bs$ 
in the effective theory.
In Sec. \ref{rpvsusy}, we present the theoretical expressions for the branching ratio of $\bs$ in the RPV mSugra.
In Sec. \ref{res}, we discuss our results.  
Finally in Sec. \ref{concl}, we present our conclusions.
%
\section{$\bs$ in the effective theory }
\label{bsbr}
The most general model independent form of the effective 
Lagrangian for the quark level transition $b \to s \mu^+ \mu^-$ that 
contributes to the decay $\bs$ has the form \cite{Grossman:1996qj,Guetta:1997fw} 
\begin{eqnarray} 
  L & = & \frac{G_F \alpha}{2 \sqrt{2} \pi} 
  \left( V_{ts}^\ast V_{tb} \right) \, 
  \left\{ 
  R_A 
  (\bar{s} \, \gamma_\mu \gamma_5 \, b) 
  (\bar{\mu} \, \gamma^\mu \gamma_5 \, \mu) 
  \right. \nonumber \\ 
  & & \left. \; \; \; \; \; \; \; \; \; \; \; \; \; \; 
  + 
  R_S 
  (\bar{s} \, \gamma_5 \, b) 
  (\bar{\mu} \, \mu) 
  + 
  R_P
  (\bar{s} \, \gamma_5 \, b) 
  (\bar{\mu} \, \gamma_5 \, \mu) 
  \right\} \; , 
  \label{eqn:heff1} 
\end{eqnarray} 
where $R_P, R_S$ and $R_A$ are the strengths of the scalar, 
pseudoscalar and axial vector operators respectively.  
$R_A$ in eq.~(\ref{eqn:heff1}) is the sum of SM and new physics contributions. 
Here we assume $R_A \simeq R_A^{SM}$.

In SM, the 
scalar and pseudoscalar couplings $R_S^{\rm SM}$ and $R_P^{\rm 
SM}$ receive contributions from the penguin diagrams with physical 
and unphysical neutral scalar exchange and are highly suppressed: 
\begin{equation} 
 R_S^{\rm SM} = R_P^{\rm SM} \propto \frac {(m_{\mu} m_b)}{m_{W}^{2}} \sim 10^{-5} \; . 
\end{equation} 
Also, $R_A^{\rm SM } = {Y(x)}/{\sin^2 \theta_W}$, where $Y(x)$ is 
the Inami-Lim function \cite{inamilim} 
\begin{equation} 
Y(x)=\frac{x}{8}\, \left[\frac{x-8}{x-1}+ \frac{3x}{(x-1)^2} \ln x 
\right] \; , 
\end{equation} 
with $x =({m_t}/{M_W})^2$. Thus, $R_A^{\rm SM }\simeq 4.3$.

The calculation of the branching ratio gives \cite{Alok:2008hh} 
\begin{equation} 
B(\bs)=a_s\left[|b_{SM}-b_P|^2\,+\,|b_S|^2\right]\; , 
  \label{blep_gen} 
\end{equation}  
where 
\begin{equation} 
a_s \equiv \frac{G_F^2 \alpha^2}{64 \pi^3} \, 
  \left| V_{ts}^\ast V_{tb}  \, \right|^2 \tau_{B_s} f_{B_s}^2 m^3_{B_s} \, 
 \sqrt{ 1 - \frac{4 m_\mu^2 }{m_{B_s}^2} } \; ,
\end{equation} 
\begin{equation} 
b_{SM}=2 \frac{m_\mu}{m_{B_s}} R_A^{SM}\;, \quad
b_{P}=\frac{m_{B_s} } {m_b + m_s} R_P\;, \quad
b_{S}=\sqrt{1 - \frac{4 m_\mu^2 }{m_{B_s}^2}}  
\frac{m_{B_s}}{m_b + m_s} R_S\;. 
\label{b-def}
\end{equation}
Here $\tau_{B_s}$ is the lifetime of $B_s$.

\section{$\bs$ in the R-parity violating mSugra}
\label{rpvsusy}

To begin with, the MSSM superpotential, written in terms of the Higgs, quark and
lepton superfields, is given by

\begin{equation}
W_{MSSM} = Y^l_{ij} L_i H_d E^c_j + Y^d_{ij} Q_i H_d D^c_j
         + Y^u_{ij} Q_i H_u U^c_j + \mu H_d H_u                                             
\end{equation}

\noindent
where $L$, $Q$ are the left-chiral lepton and quark superfields 
respectively and $E$, $U$, $D$ are the right-chiral lepton, up-type and down-type quark superfields respectively and $H_d$ and $H_u$ are respectively the Higgs superfileds. $i$, $j$  are family indices. $Y$'s denote the Yukawa matrices and $\mu$ is the Higgsino mass parameter. 
Now, if $R=(-)^{(3B+L+2S)}$ is not conserved, the superpotential
admits of the following additional terms~\cite{rp}:

\begin{equation}
W_{R_p{\!\!\!\!\!\!/\ }} = \lambda_{ijk} L_i L_j E^c_k
              + \lambda^{'}_{ijk} L_i Q_j D^c_k
              + \lambda^{''}_{ijk} \bar{U}_i D^c_j \bar{D}_k
              + \epsilon_{i} L_{i} H_u
\end{equation}

\noindent
where the terms proportional to $\lambda_{ijk}$, $\lambda_{ijk}^{'}$
and $\epsilon_i$ violate lepton number, and those proportional to
$\lambda_{ijk}^{''}$ violate baryon number. The constants
$\lambda_{ijk}$ ($\lambda_{ijk}^{''}$) are antisymmetric in the first
(last) two indices.

In case of  exact SUSY, masses of superparticles are the same as their SM counterpart. Now, since no superparticle has been observed till date in experiments, which means SUSY must be broken in masses. SUSY breaking in masses can be achieved by the introduction of soft mass-terms corresponding to each superparticles. Thus in SUSY we have a large number of free parameters ($\sim$ 150). In a mSugra framework,  where same spin particles share a common (soft) mass at a high scale ($\Lambda \sim 10^{16} GeV$), the whole SUSY spectrum can be described in terms of five (5) free parameters at 
high scale $\Lambda$. These are 
as follows 
\begin{itemize}
\item A universal scalar mass, $m_0$ 
\item A universal fermion mass, $m_{1/2}$
\item A universal trilinear coupling, $A$ 
\item $\tan\beta=<H_d>/<H_u>$, the ratio of vaccum expectation value (vev) of the down and up type Higgses, and 
\item $sgn(\mu)$, sign of the Higgsino mass parameter.
\end{itemize}

We now compute $R_S$ and $R_P$, the Wilson coefficients of the scalar and pseudoscalar 
operators, in the $b\to s \mu^+ \mu^-$ transition, within the context of RPV mSugra. We focus on
the large $\tan\beta $ scenario ($\tan\beta > 30$). There are three contributions to the $R_S$ and $R_P$: (i) Contribution from the Higgs sector, 
(ii) Contribution from the RPC sector and, (iii) Contribution from the RPV sector. 

\subsection{Contribution from the Higgs sector}

In SUSY, up-type quarks couple to one Higgs doublet ($H_u$) while the 
down-type quarks couple to the other Higgs doublet ($H_d$) which is the same as the type II-two Higgs doublet model (2HDM).

Retaining only leading terms in $\tan\b$ (which is a valid approximation for $\tan\beta > 30$), the Higgs contribution to 
$R_S$ and $R_P$ is given by \cite{Bobeth:2001sq}
\be
R_S^{\rm 2HDM} = -R_P^{\rm 2HDM} = \frac{m_b\,m_l\,\tan^2\beta}{4M_W^2\sin^2\theta_W}
\frac{\ln r}{1-r},\quad r=\frac{m_{H^{\pm}}^2}{m_t^2}.
\label{constrained:2HDM:res}
\ee

\subsection{Contribution from the RPC sector}

We consider a scenario with minimal flavour violation, i.e. we assume 
flavour-diagonal sfermion mass matrices, the contributing 
SUSY diagrams, in addition to those contributing to 2HDM, 
consist only of the two chargino states.
The RPC contribution to 
$R_S$ and $R_P$ is given by \cite{Bobeth:2001sq}
\be
R_S^{RPC}=R_{S}^{\Box}+R_{S}^{\Peng}+R_S^{\Count}\;,
\ee
\be
R_P^{RPC}=R_{P}^{\Box}+R_{P}^{\Peng}+R_P^{\Count}\;,
\ee
where
\bea\label{box}
R_{S,P}^{\Box} = &\mp& \frac{m_b m_l\tan^2\beta}
{2M_W}\sum_{i,j=1}^{2}\sum_{a=1}^{6}\sum_{k,m,n=1}^{3}
\frac{1}{m_{\tilde{\chi}_i^{\pm}}^2}
\Bigg\{
(R^\dagger_{\sneutrino})_{lk}(R_{\sneutrino})_{kl}
     (\Gamma^{U_L})_{am}U_{j2}\G^a_{imn}\nnu\\
   &\times&\Bigg(y_{ai} U_{j2}^{\ast}
V_{i1}^{\ast}\pm\frac{m_{\tilde{\chi}_j^{\pm}}}{m_{\tilde{\chi}_i^{\pm}}}
     U_{i2}V_{j1}\Bigg)D_1(x_{ki},y_{ai},z_{ji})\Bigg\},
\eea
\bea\label{peng}
   R_{S,P}^{\Peng}  = &\pm&\frac{m_b m_l\tan^2\beta}{M_W^2(m_{H^{\pm}}^2-M_W^2)}
\sum_{i,j=1}^{2}\sum_{a,b=1}^{6}\sum_{k,m,n=1}^{3}
     \G^a_{imn}(\Gamma^{U_L})_{bm}U_{j2}\nnu \\
   &\times&\Bigg\{M_W\Bigg(y_{aj}
     U_{j2}^{\ast}V_{i1}^{\ast}\pm\frac{m_{
     \tilde{\chi}_i^{\pm}}}{m_{\tilde{\chi}_j^{\pm}}}U_{i2}V_{j1}\Bigg)D_2(y_{aj},z_{ij})\delta_{ab}\delta_{km}\nnu\\
&-&\frac{ (M_U)_{kk}}{\sqrt{2}m_{\tilde{\chi}_i^{\pm}}}[\m^*(\Gamma^{U_R})_{ak}(\Gamma^{U_L\dagger})_{kb}
     \pm\m(\Gamma^{U_L})_{ak}(\Gamma^{U_R\dagger})_{kb}]
     D_2(y_{ai},y_{bi})\delta_{ij}\Bigg\},
\eea
\bea\label{count}
R_{S,P}^{\Count} &=&\mp
\frac{m_b m_l\tan^3\beta}{\sqrt{2}M_W^2
(m_{H^{\pm}}^2-M_W^2)}\sum_{i=1}^{2}\sum_{a=1}^{6}\sum_{m,n=1}^{3}
 [m_{\tilde{\chi}_i^{\pm}} D_3(y_{ai})U_{i2}(\Gamma^{U_L})_{am}\G^a_{imn}],
\eea
where  
\be
M_U\equiv \diag(m_u, m_c, m_t),
\ee
\be\label{gamma}
\G^a_{imn}= \frac{1}{2\sqrt{2}\sin^2\theta_W}
[\sqrt{2}M_W V_{i1}(\Gamma^{U_L\dagger})_{na}-(M_U)_{nn}V_{i2}
     (\Gamma^{U_R\dagger})_{na}]\la_{mn}\;.
\ee
The mass ratios are defined as 
\be
   x_{ki}=\frac{m_{\sneutrino_k}^2}{m_{\tilde{\chi}_i^{\pm}}^2},\quad
   y_{ai}=\frac{m_{\upsquark_a}^2}{m_{\tilde{\chi}_i^{\pm}}^2},\quad
   z_{ij}=\frac{m_{\tilde{\chi}_i^{\pm}}^2}{m_{\tilde{\chi}_j^{\pm}}^2}\;,
\ee
with $\sneutrino_k$, $\upsquark_a$, and $\chargino_i$ denoting sneutrinos, 
up-type squarks, and charginos. 
The ratio of CKM factors 
$\la_{mn}\equiv V_{mb}^{}V_{ns}^{\ast}/V_{tb}^{}V_{ts}^{\ast}$, and the 
functions $D_{1,2,3}$ are listed in Appendix A of ref. \cite{Bobeth:2001sq}.

\subsection{Contribution from the RPV sector }
The RPV contribution to 
$R_S$ and $R_P$ is given by \cite{Mir:2008zz,Dreiner:2006gu}
\be
R_S^{RPV}=\frac{1}{4}\left(B^{bq'}_{\beta \beta}-C^{bq'}_{\beta \beta}\right)\;,
\ee
\be
R_P^{RPV}=\frac{1}{4}\left(B^{bq'}_{\beta \beta}+C^{bq'}_{\beta \beta}\right)\;,
\ee
where
\begin{equation}
B_{\beta \beta }^{bq'}=\frac{2\sqrt{2}\pi}{G_{F}\alpha }\underset{i=1}{\overset{3%
}{\sum }}\frac{1}{V_{tb}V_{tq'}^{\ast }}\frac{2\lambda _{i\beta \beta }^{\ast
}\lambda _{iq'3}^{\prime }}{m_{\widetilde{\nu }_{iL}}^{2}}
\end{equation}%
\begin{equation}
C_{\beta \beta }^{bq'}=\frac{2\sqrt{2}\pi}{G_{F}\alpha }\underset{i=1}{\overset{3%
}{\sum }}\frac{1}{V_{tb}V_{tq'}^{\ast }}\frac{2\lambda _{i\beta \beta
}\lambda _{i3q'}^{\prime \ast }}{m_{\widetilde{\nu }_{iL}}^{2}}
\end{equation}%
Here $q'=s$ and $\beta=\mu$.

\begin{table}[t]
\begin{center}
\begin{tabular}{|l|}
\hline
$G_F = 1.166 \times 10^{-5} \; \gev^{-2}$ \\ $\alpha = 1.0/137.0$ \\
$\tau_{B_s} = 1.45 \times 10^{-12}\; s$ \\ $m_{B_s}=5.366 \; \gev$ \\ 
$m_{\mu}=0.105 \;\gev$  \\ $m_b=4.20\; \gev $ \\
$m_s=0.100\; \gev $\\  $m_t=172.3\; \gev $ \\
$m_W=80.403\; \gev $\\ $V_{tb}= 1.0 $  \\
$|V_{ts}|= (40.6 \pm 2.7) \times 10^{-3}$ \\$f_{B_s}=(259\pm 27)\;$ MeV\;\cite{Mackenzie:2006un}\\ 
\hline
\end{tabular}
\caption{\small\sf Numerical inputs used in our 
  analysis. Unless explicitly specified, they are taken from the 
  Review of Particle Physics~\cite{PDG}. }
\label{table1}
\end{center}
\end{table}
\begin{table}[t]
\begin{center}
\begin{tabular}{|c|c|c|}
\hline
Parameter & Range\\ \hline
$m_0$ & $100 - 2000$ \gev\\
$m_{1/2}$ & $100 - 2000$ \gev\\
$A$       &  $-3000 - 3000$ \gev\\
$\tan\beta$& $30 - 80$\\
$sgn(\mu)$ & $+$, $-$\\
\hline
\end{tabular}
\caption{\small\sf Ranges of various Sugra parameters used in our analysis.}
\label{sparam}
\end{center}
\end{table}

\section{Numerical Analysis and Results}
\label{res}

We have already seen in section 2, the low energy 
SUSY mass spectrum can be fully obtained by specifying the mSugra parameter set 
\{$m_0, m_{1/2}, A, sgn(\mu)\, {\rm and} \tan\beta$\} at the high scale once R-parity is conserved. 
In order to generate low energy SUSY mass spectrum, we use {\bf 
Suspect2.0} \cite{suspect}. Ranges of various High scale 
mSugra parameters used our analysis are summarized in Table \ref{sparam}. Finally for the sake of simplicity, in case of RPV-Sugra, 
we assume $\lambda_{ijk} = \lambda$ and $\lambda^{'}_{ijk} = \lambda^{'}$ and $\lambda^{''}_{ijk} = \lambda^{''}$ at the electroweak scale.

While scanning the parameter space we have taken into account  upper bound on the supersymmetric contribution to the $\rho$-parameter~\cite{Barbieri:1983wy,Lim:1983re,Eliasson:1984yu,Drees:1990dx,Djouadi:1996pa,Djouadi:1998sq}, anomalous magnetic moment of the muons~\cite{Davier:2003pw,Hagiwara:2003da,de Troconiz:2004tr,Passera:2004bj}, 
bounds on $b\to s \gamma$ branching fraction~\cite{PDG,Kagan:1998ym,Gambino:2001au,Gambino:2000fz,Gambino:2003zm} 
and LEP bounds on the Lightest neutral Higgs mass 
and superparticle masses~\cite{PDG,lep}.

Further constraints has been employed in case of RPC Sugra in order to 
avoid charged lightest superparticle (LSP) as here, unlike RPV case, the 
LSP has be to stable as a consequence of R-parity conservation. Finally, 
in order to respect existing bounds on the products in RPV couplings 
from earlier analysis 
\cite{rp,Bhattacharyya:1996nj,Dreiner:1997uz,Allanach:1999ic}, we use 
$|\lambda . \lambda^{'}| < 10^{-5}$. It is to be noted that the products 
involving $\lambda^{''}$ will not affect our analysis. This is simply because 
pure baryon number violating terms does not contribute to the $\bs$ 
branching ratio.

\subsection{Predictions for $\bs$ branching ratio}

\begin{figure}
\centerline{
{\label{fig:mhbrv0}\includegraphics[angle=-90, width=0.5\textwidth]{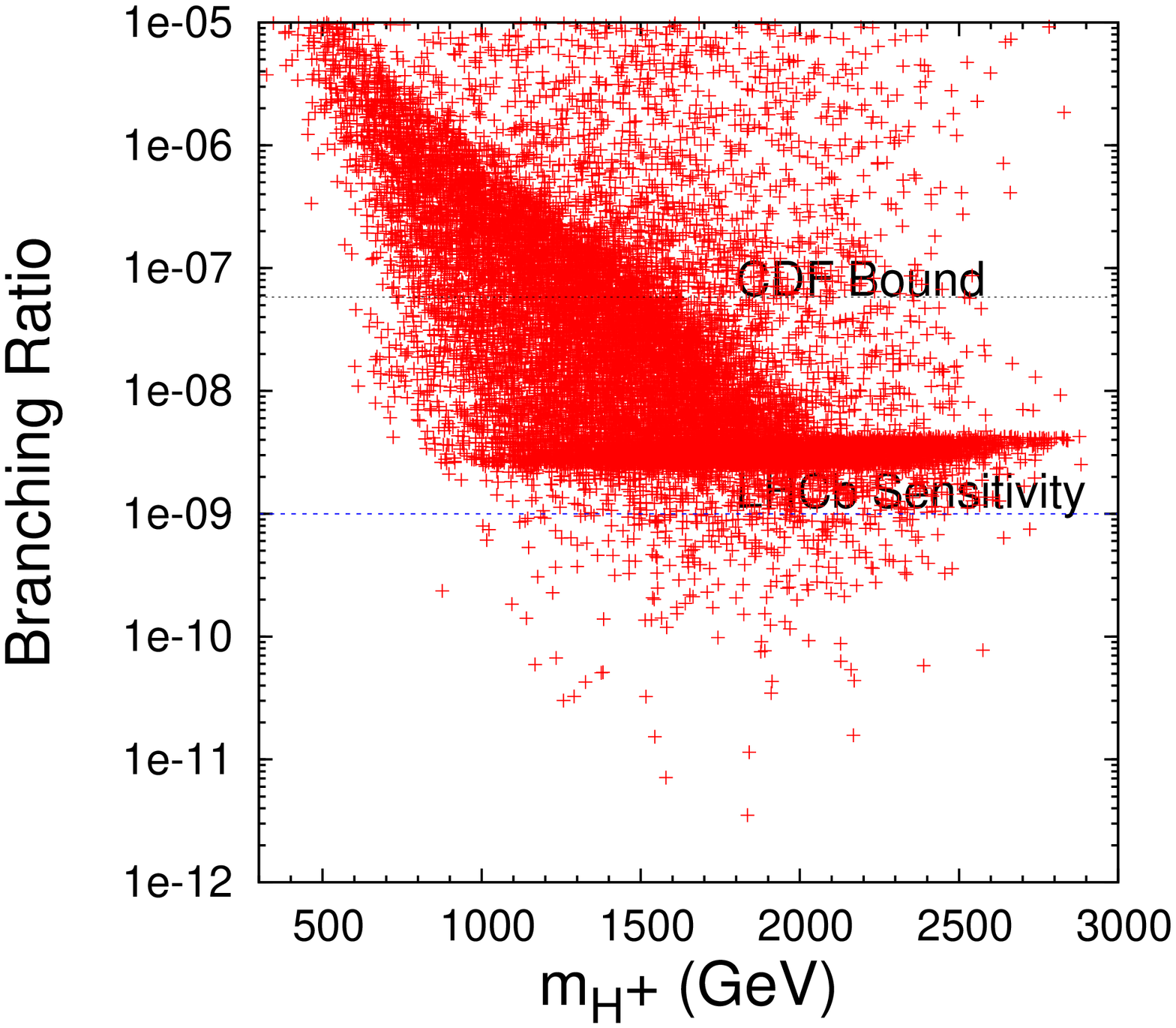}}  
{\label{fig:mhbrsv0}\includegraphics[angle=-90, width=0.5\textwidth]{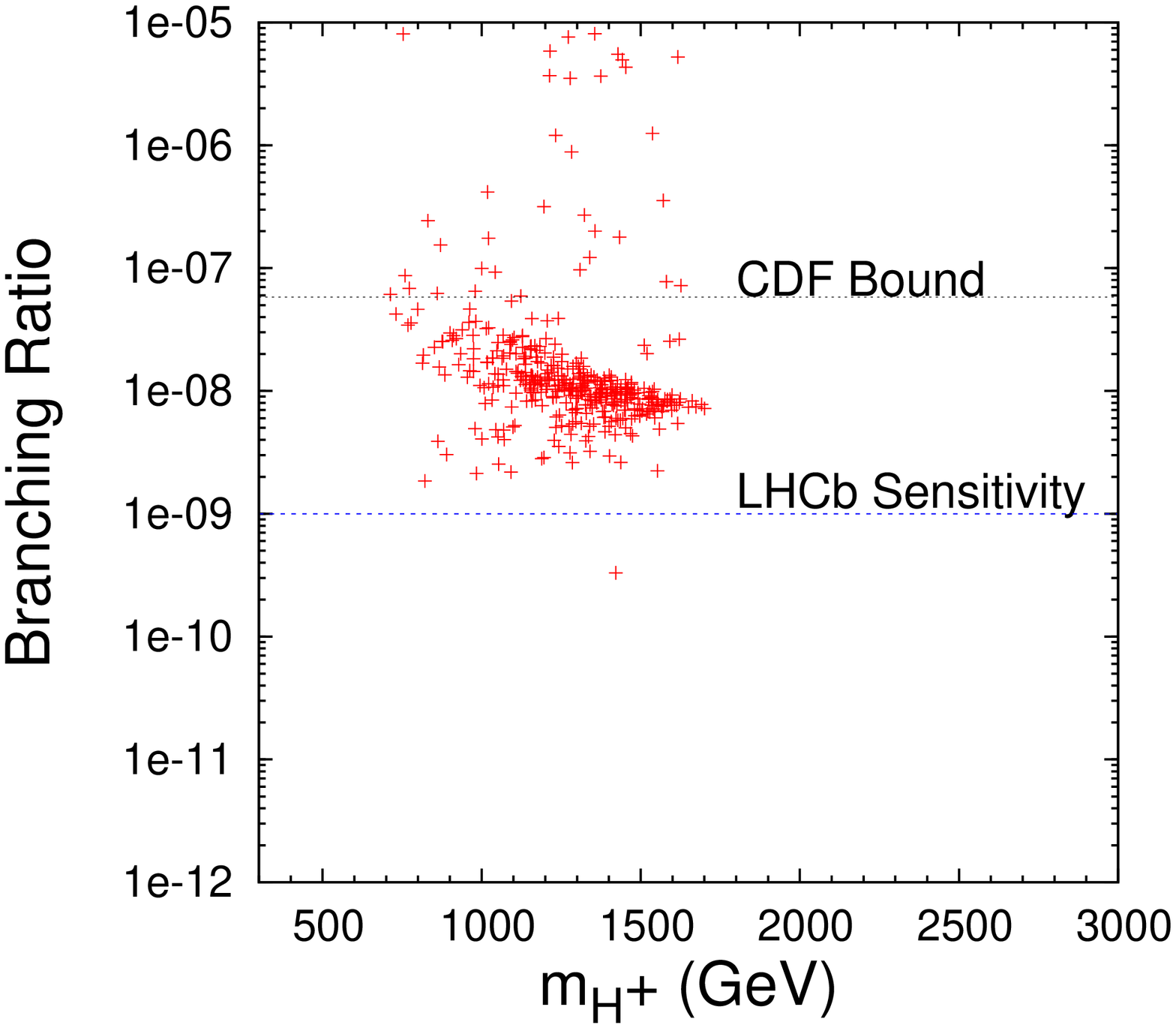}} 
}
\caption{\small\sf RPV branching ratio for $\bs$ vs $m_{H^\pm}$. The left panel of the Figure corresponds to 
$\mu>0$ whereas the right panel corresponds to $\mu<0$.}
\label{fig:rpv5}
\end{figure}
\begin{figure}
\centerline{
{\label{fig:c0c1v0}\includegraphics[angle=-90, width=0.5\textwidth]{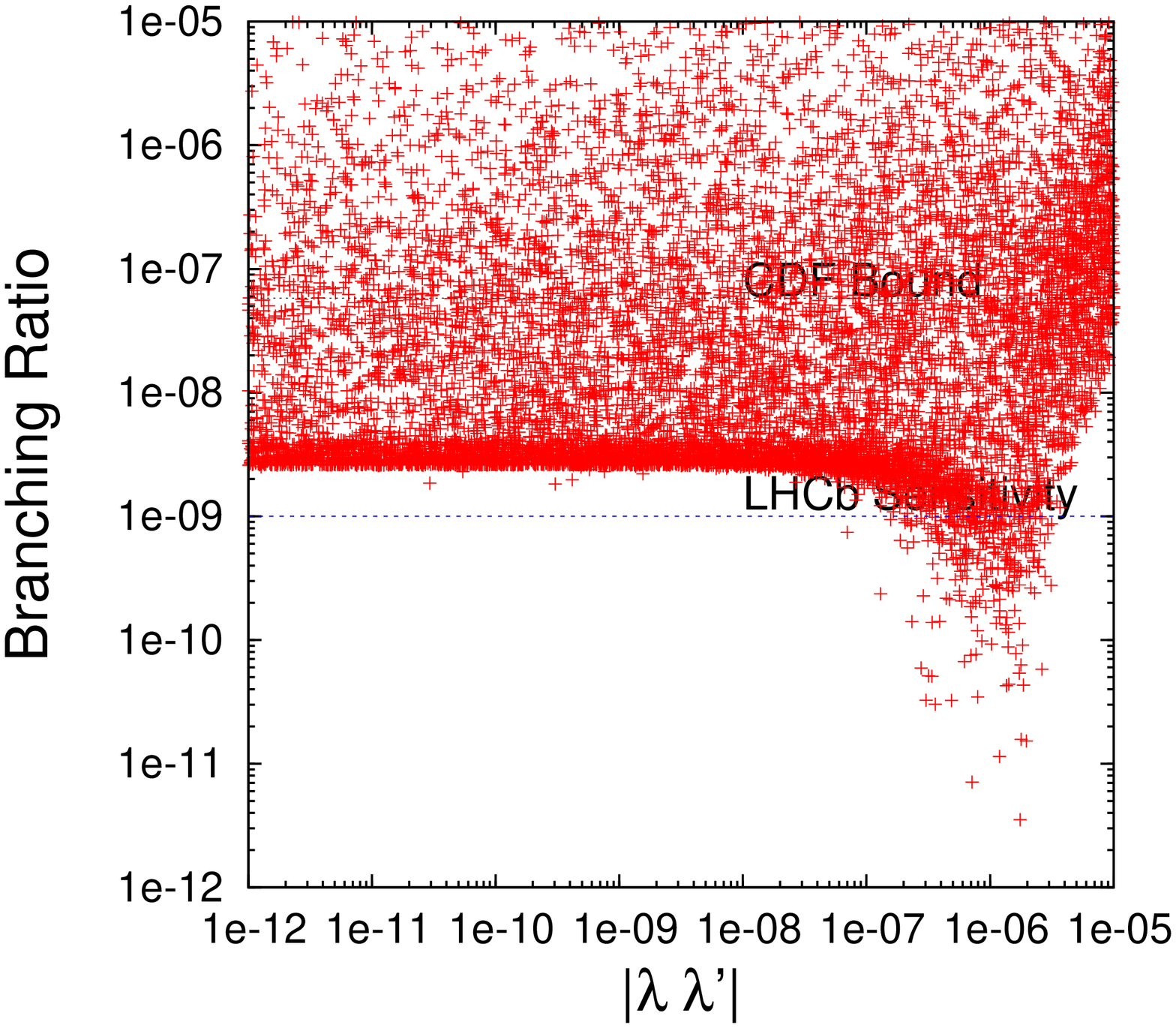}}  
{\label{fig:c0c1sv0}\includegraphics[angle=-90, width=0.5\textwidth]{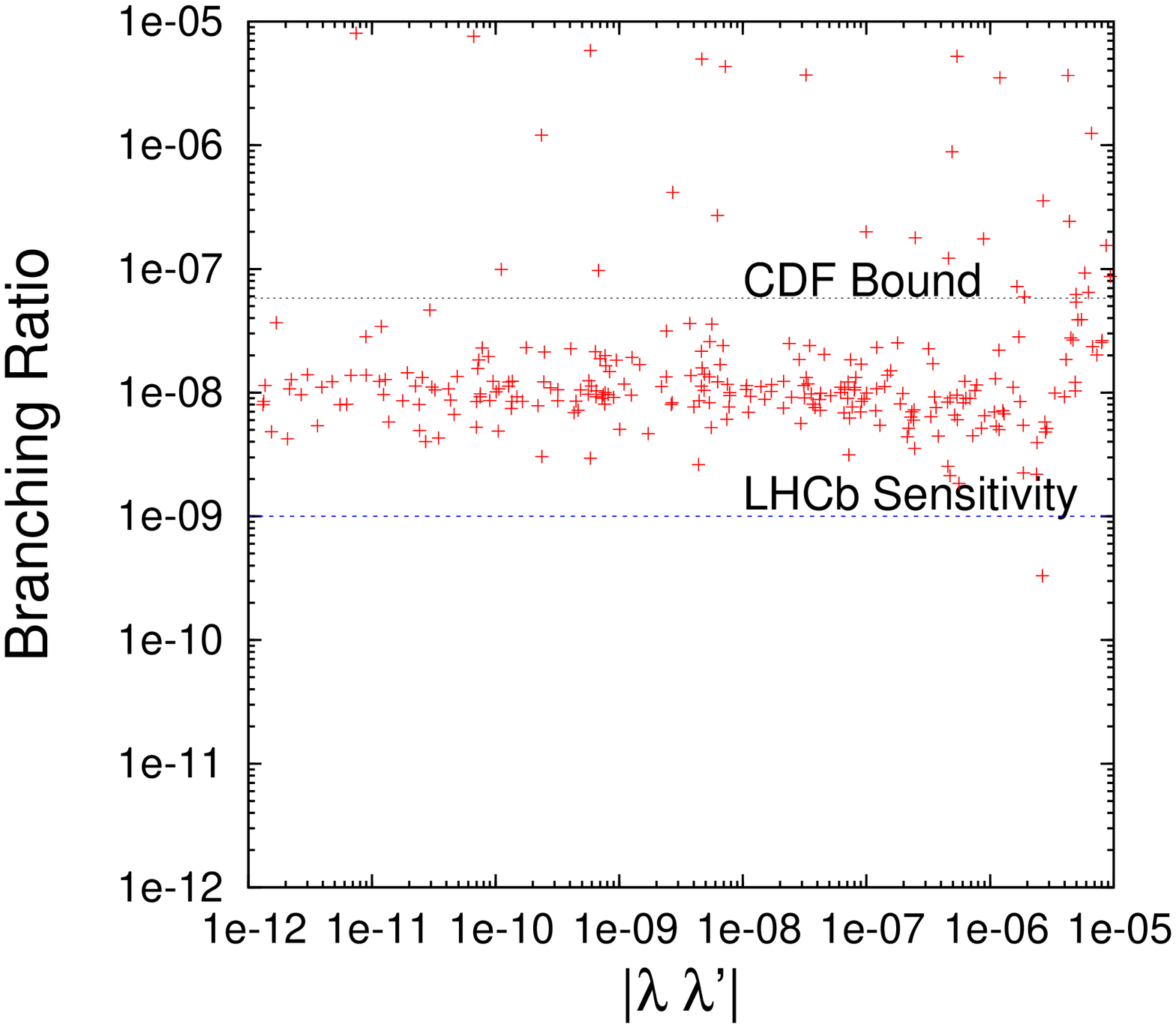}}
}
\caption{\small\sf RPV branching ratio for $\bs$ vs $|\lambda\lambda^\prime|$. The left panel of the Figure corresponds to  
$\mu>0$ whereas the right panel corresponds to $\mu<0$.}
\label{fig:rpv6}
\end{figure}

The possible values of $\br$ in the RPV mSugra as a function of $m_{H^\pm}$ and $|\lambda\lambda^\prime|$ are shown in Figure \ref{fig:rpv5} and \ref{fig:rpv6} respectively. 
It is obvious from the Figure that the branching ratio of 
$\bs$ in the RPC mSugra can be enhanced by more than an order of magnitude above its 
SM expectation. In that case $\bs$ can even be observed at the tevatron. 
On the other hand, $\br$ can also be suppressed to a value as low 
as $2 \times  10^{-12} $ for $\mu >0$ ($3 \times  10^{-10} $ for $\mu<0$ ) 
which is well below the present LHCb sensitivity for $\bs$. In such a situation $\bs$ can be invisible to the LHC. 
The possibility of invisibility of $\bs$ at the LHC due to new physics scalar/pseudoscalar 
operators was first pointed out in the ref. \cite{Alok:2008hh} whereas in \cite{Dedes:2008iw} 
it was shown that in the minimal supersymmetric standard model, 
$\br$ can go well below the SM predictions in the low $\tan\beta$ regime ($\tan\beta< 10$).

Thus we see that even in the large $\tan \beta$ regime, the lowest 
value of $\br$ in the RPV mSugra can go several orders of magnitude 
below the present LHCb sensitivity. Hence $\bs$ can even be invisible to the LHC.

\subsection{Constraints on the RPV mSugra parameter space from the upper bound on $\br$ }

We now study the constraints imposed on the RPV mSugra parameter space from the upper bound on $\bs$. 
In the limit $|\lambda . \lambda'| \rightarrow 0$, which is the case of RPC mSugra, we find our results consistent with previous works in literature \cite{Babu:1999hn, Chankowski:2000ng, Dedes:2001fv, Buras:2002vd, Ellis:2007ss}. Hence we discuss  constraints on RPV mSugra parameter space only. 
\begin{figure}
\centerline{
\subfloat[$5.8\times10^{-8}$]{\label{fig:msmfv0}\includegraphics[angle=-90, width=0.35\textwidth]{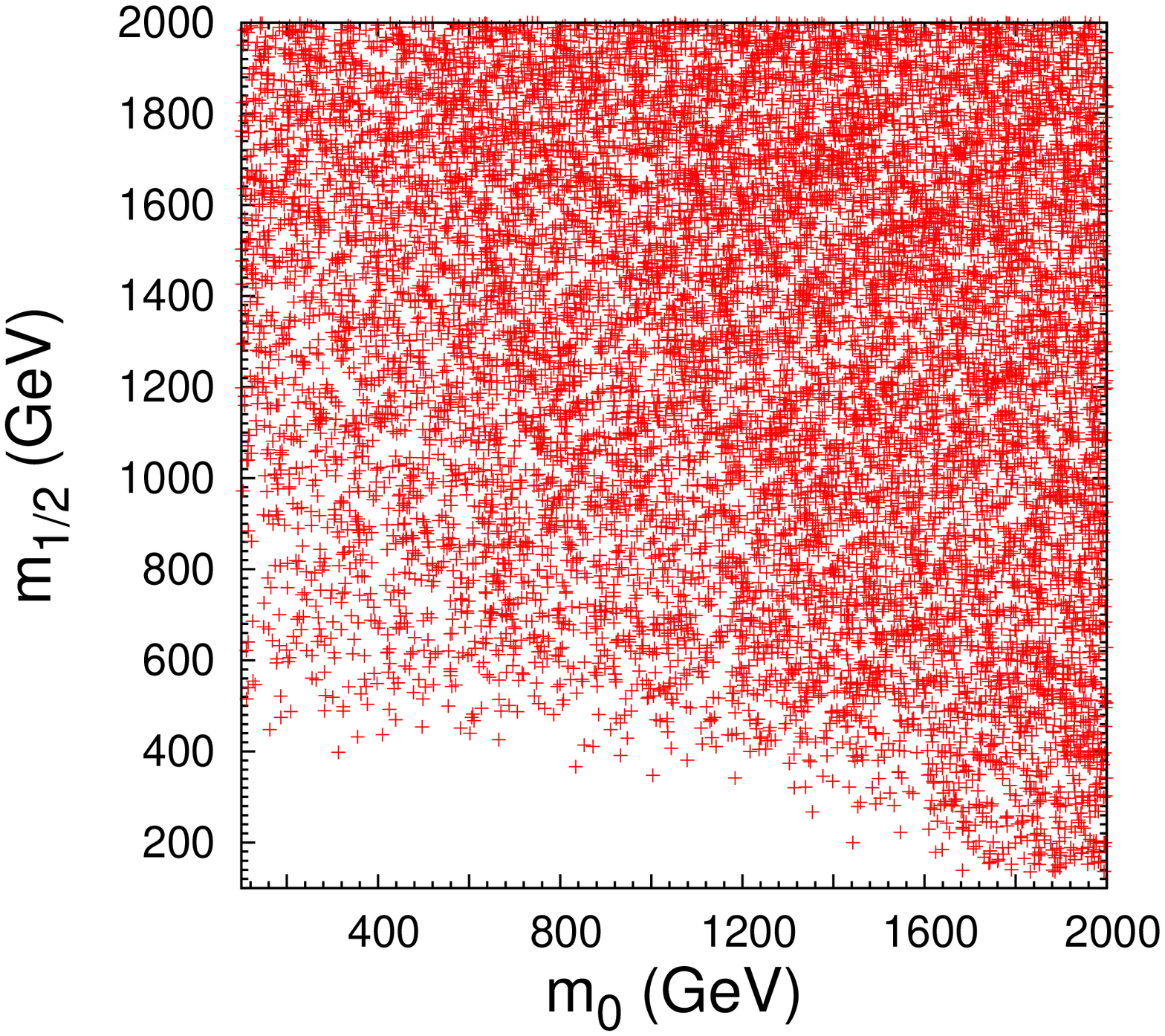}}  
\subfloat[$5.0\times10^{-9}$]{\label{fig:msmfv1}\includegraphics[angle=-90, width=0.35\textwidth]{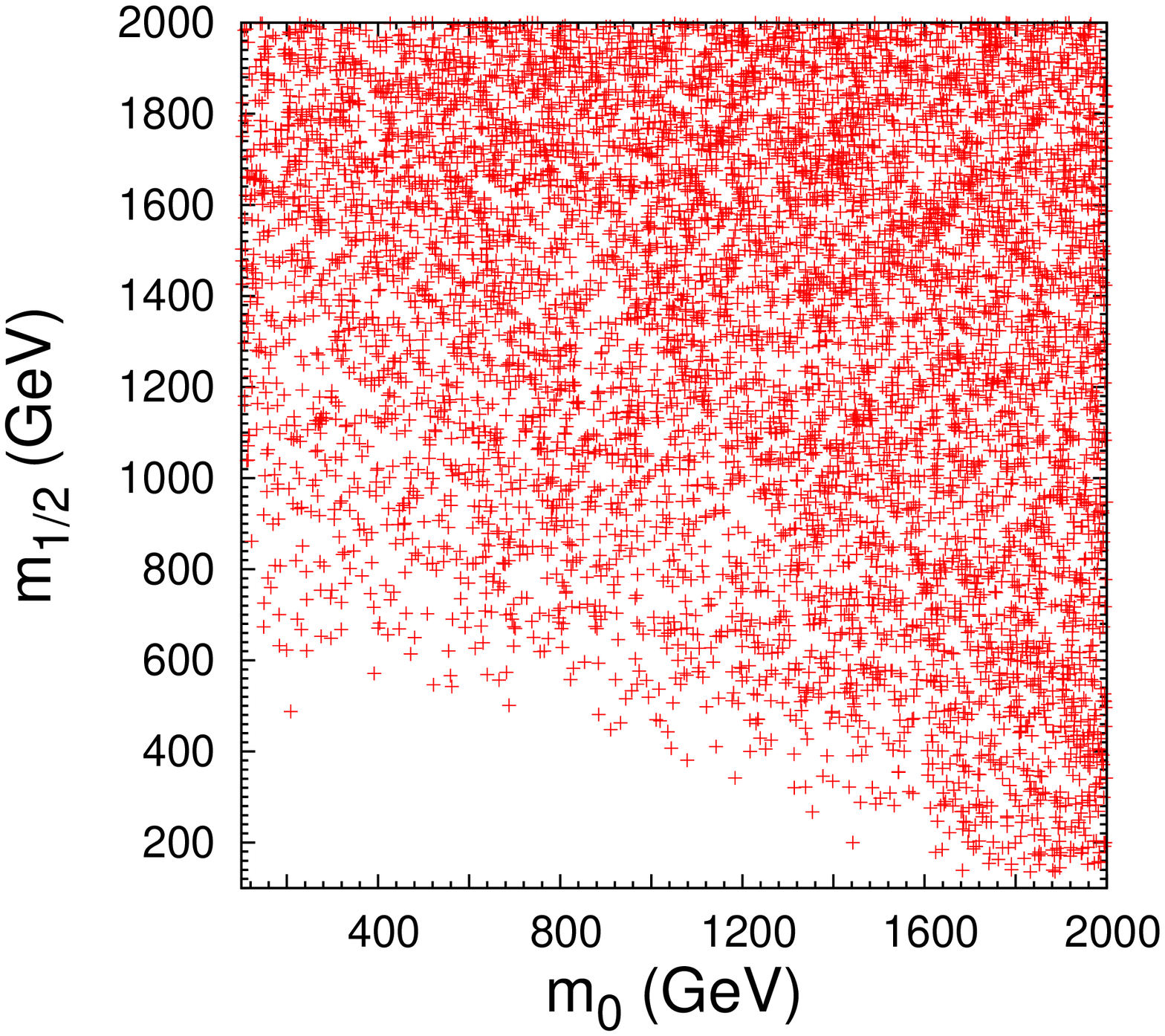}}
\subfloat[$1.0\times10^{-9}$]{\label{fig:msmfv2}\includegraphics[angle=-90, width=0.35\textwidth]{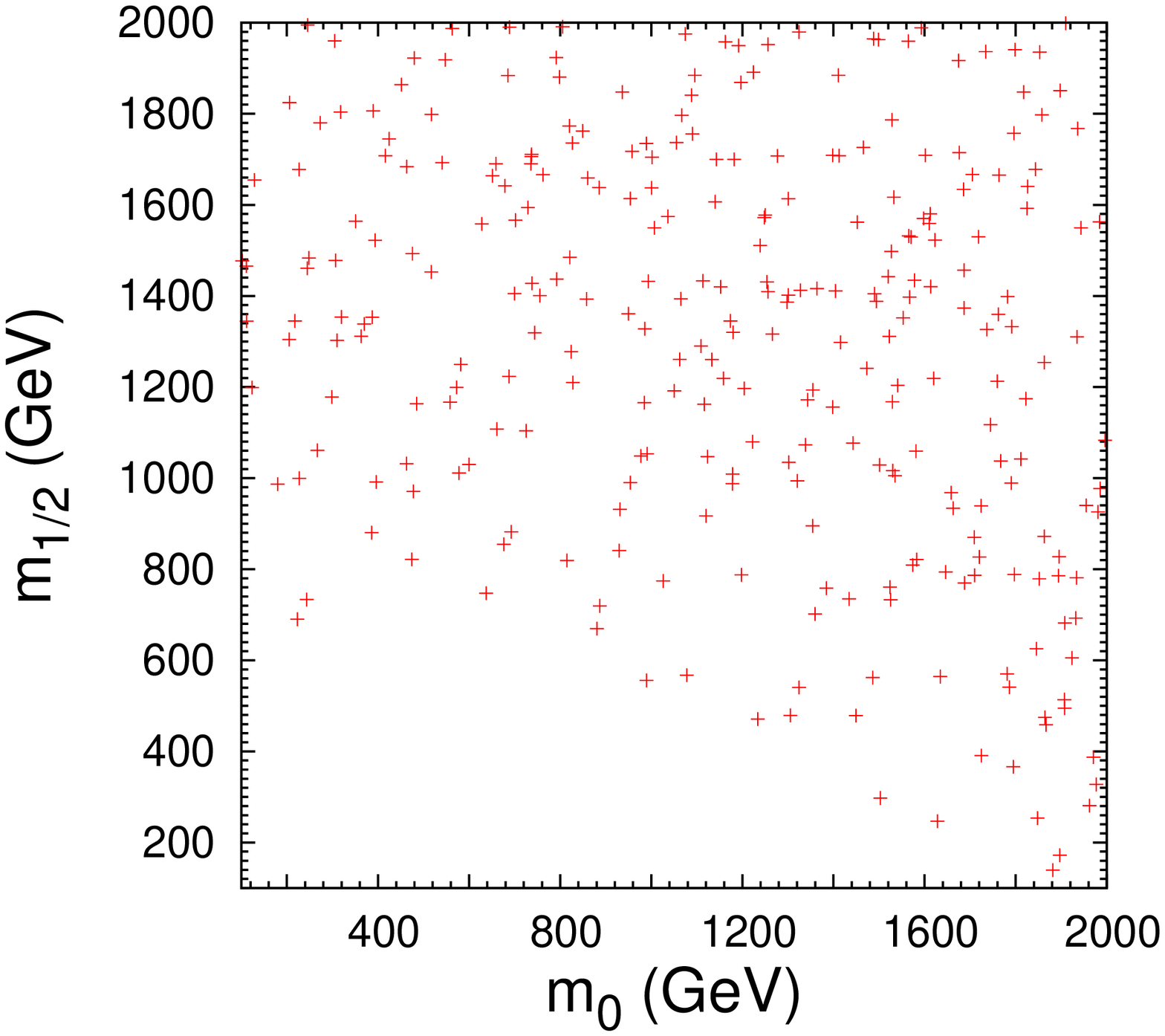}}
}
\caption{\small\sf RPV $(m_0-m_{1/2})$ plane for $\mu>0$ for $\br=(5.8,\,0.50,\,0.10) \times 10^{-8}$.}
\label{fig:rpv1}
\end{figure}

Figure \ref{fig:rpv1} shows the allowed RPV $(m_0,m_{1/2})$ plane for $\mu>0$ for several values of 
the branching ratio of $\bs$. The present upper bound on the branching ratio of $\bs$ rules out some parameter space for
$m_{1/2} \lesssim 400\, \gev$. The situation remains almost the same if the upper bound on $\br$ is brought down to $5.0 \times 10^{-9}$. 
However if the upper bound on $\br$ is  as low as $1.0 \times 10^{-9}$, 
which is the LHCb sensitivity for  $\br$, a large $(m_0,m_{1/2})$ parameter space is ruled out.
In this case all parameter space in the region
$m_0 \lesssim 900\, \gev$ and $m_{1/2} \lesssim 600\, \gev$ is ruled out. 

\begin{figure}
\centerline{
\subfloat[$5.8\times10^{-8}$]{\label{fig:msmfsv0}\includegraphics[angle=-90, width=0.35\textwidth]{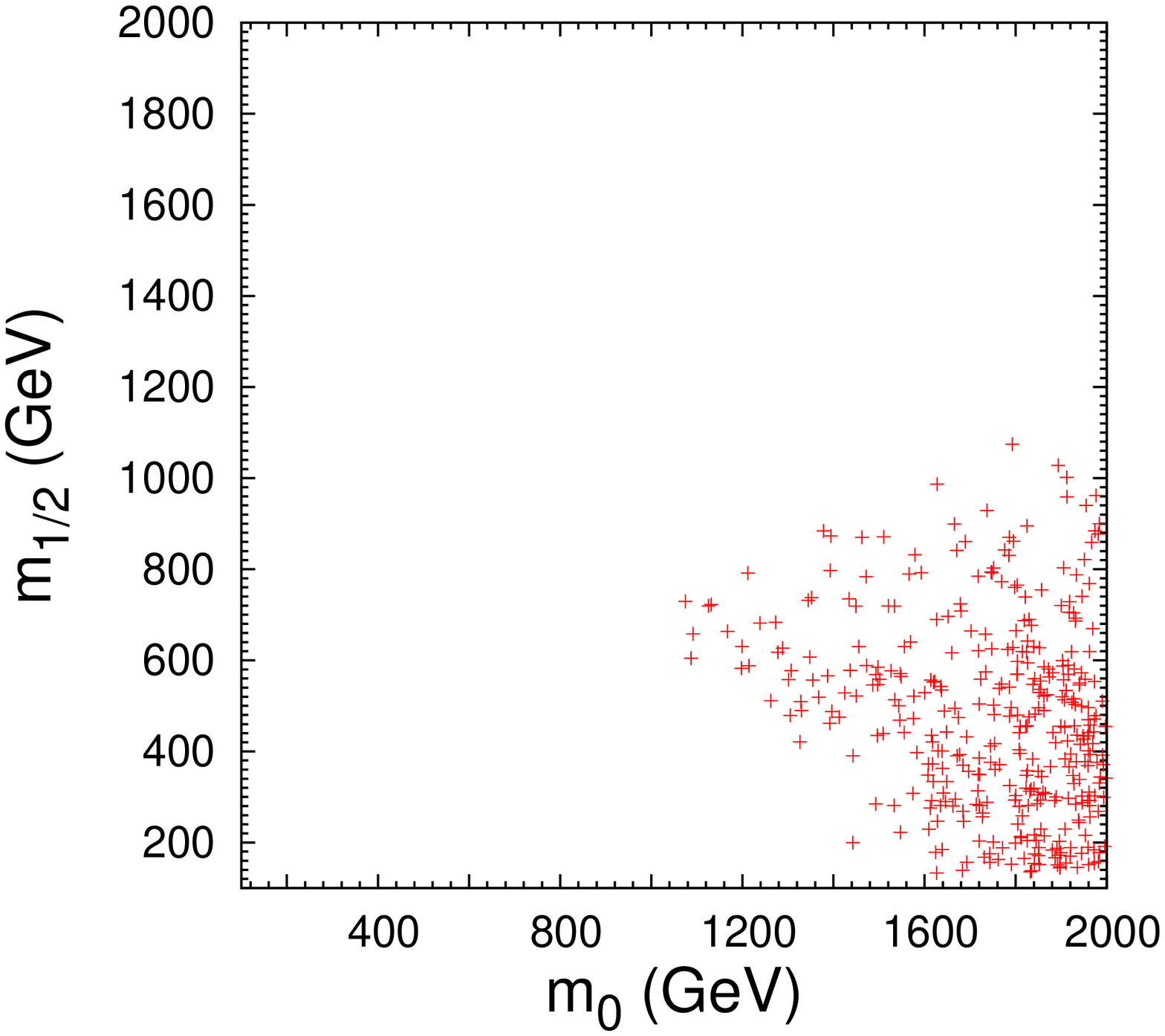}}  
\subfloat[$1.0\times10^{-8}$]{\label{fig:msmfsv1}\includegraphics[angle=-90, width=0.35\textwidth]{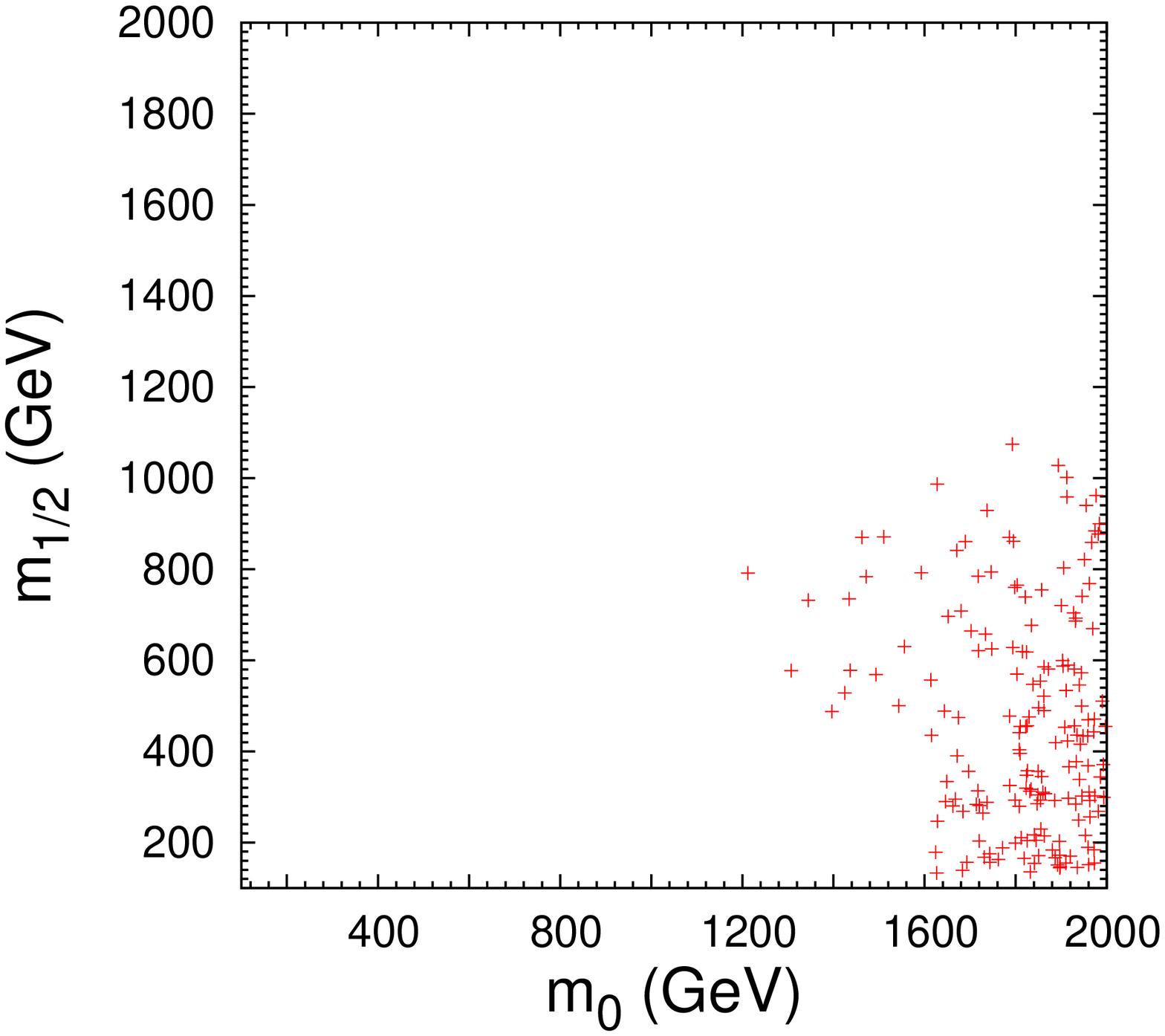}}
\subfloat[$5.0\times10^{-9}$]{\label{fig:msmfsv2}\includegraphics[angle=-90, width=0.35\textwidth]{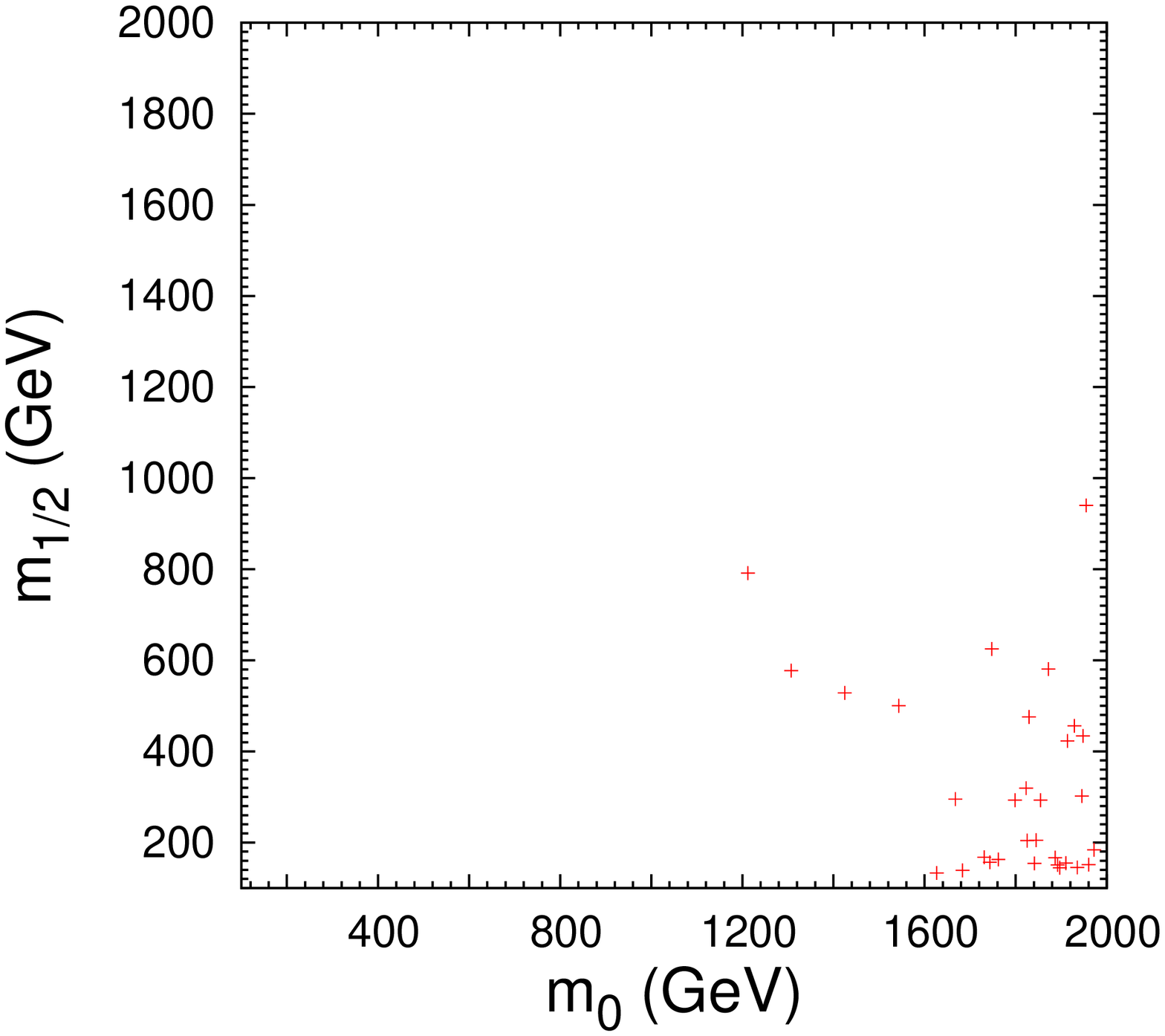}}
}
\caption{\small\sf RPV $(m_0-m_{1/2})$ plane for $\mu<0$ for $\br=(5.8,\,1.0,\,0.50) \times 10^{-8}$.}
\label{fig:rpvs1}
\end{figure}
The constraints on the RPV plane for $\mu<0$ is shown in Figure \ref{fig:rpvs1}. 
We see that in this case the upper bound on the branching ratio of $\bs$ puts more 
stringent constraint on $(m_0,m_{1/2})$ plane as compared to the case when $\mu>0$.
It can be seen from the Figure \ref{fig:msmfsv0} that the present upper bound on $\br$ rules out all parameter space for 
$m_0 \lesssim 1000\, \gev$. All parameter space above $m_{1/2} \gtrsim 1100\, \gev$ is also ruled out. 
For the upper bound on $\br$ close to the SM prediction, almost all parameter is ruled out. 

\begin{figure}
\centerline{
\subfloat[$5.8\times10^{-8}$]{\label{fig:asmsv0}\includegraphics[angle=-90, width=0.35\textwidth]{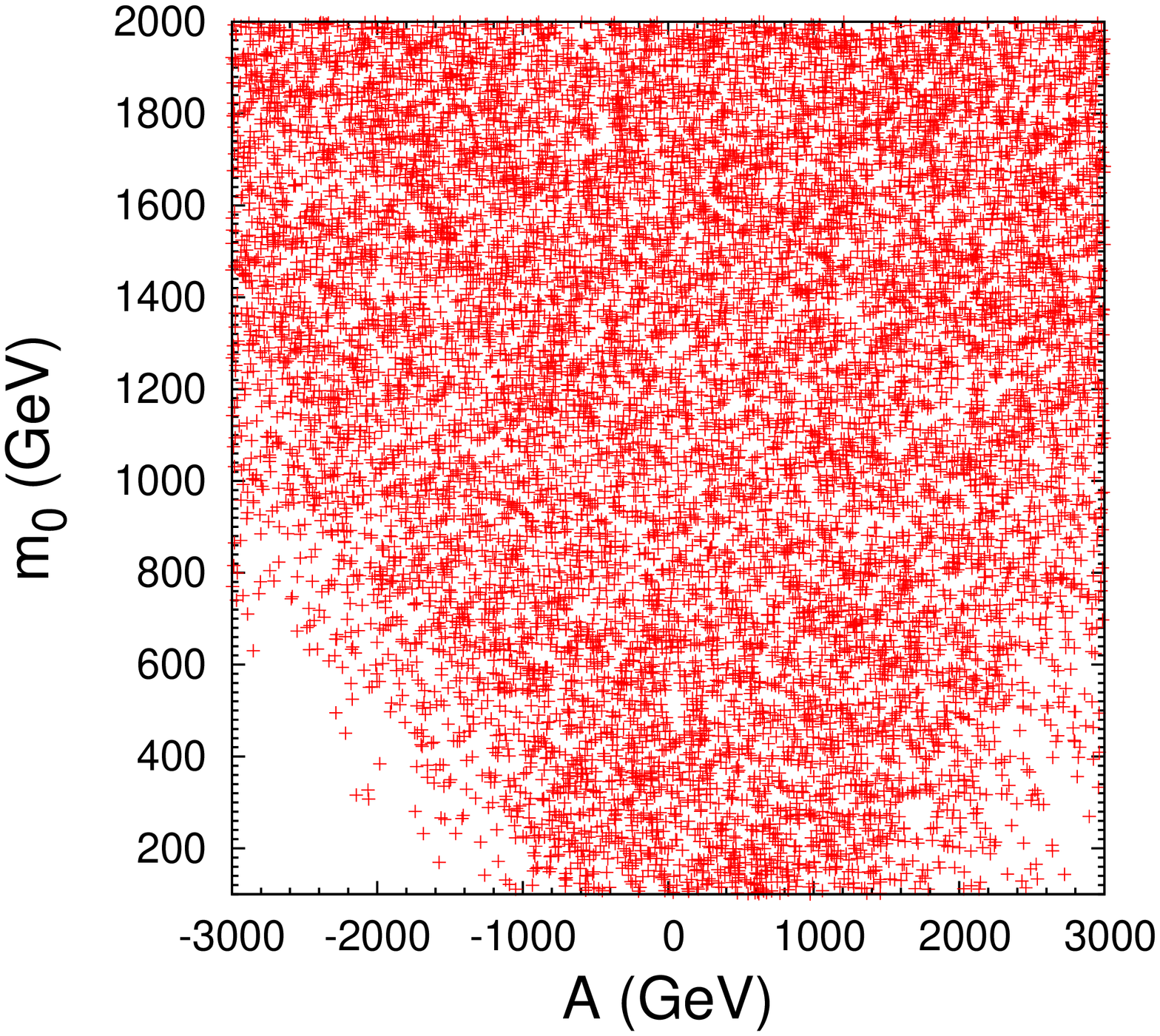}}  
\subfloat[$5.0\times10^{-9}$]{\label{fig:asmsv1}\includegraphics[angle=-90, width=0.35\textwidth]{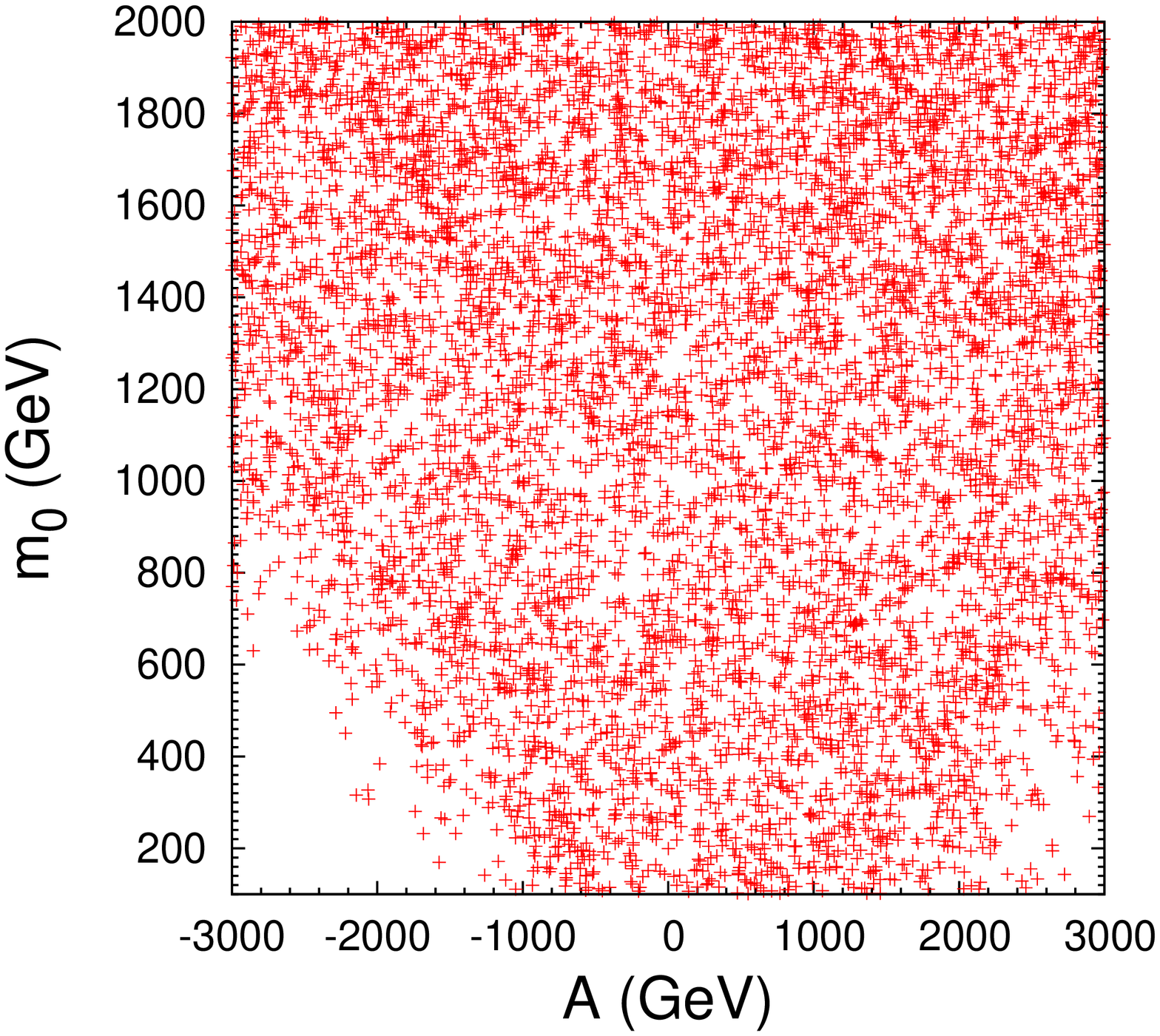}}
\subfloat[$1.0\times10^{-9}$]{\label{fig:asmsv2}\includegraphics[angle=-90, width=0.35\textwidth]{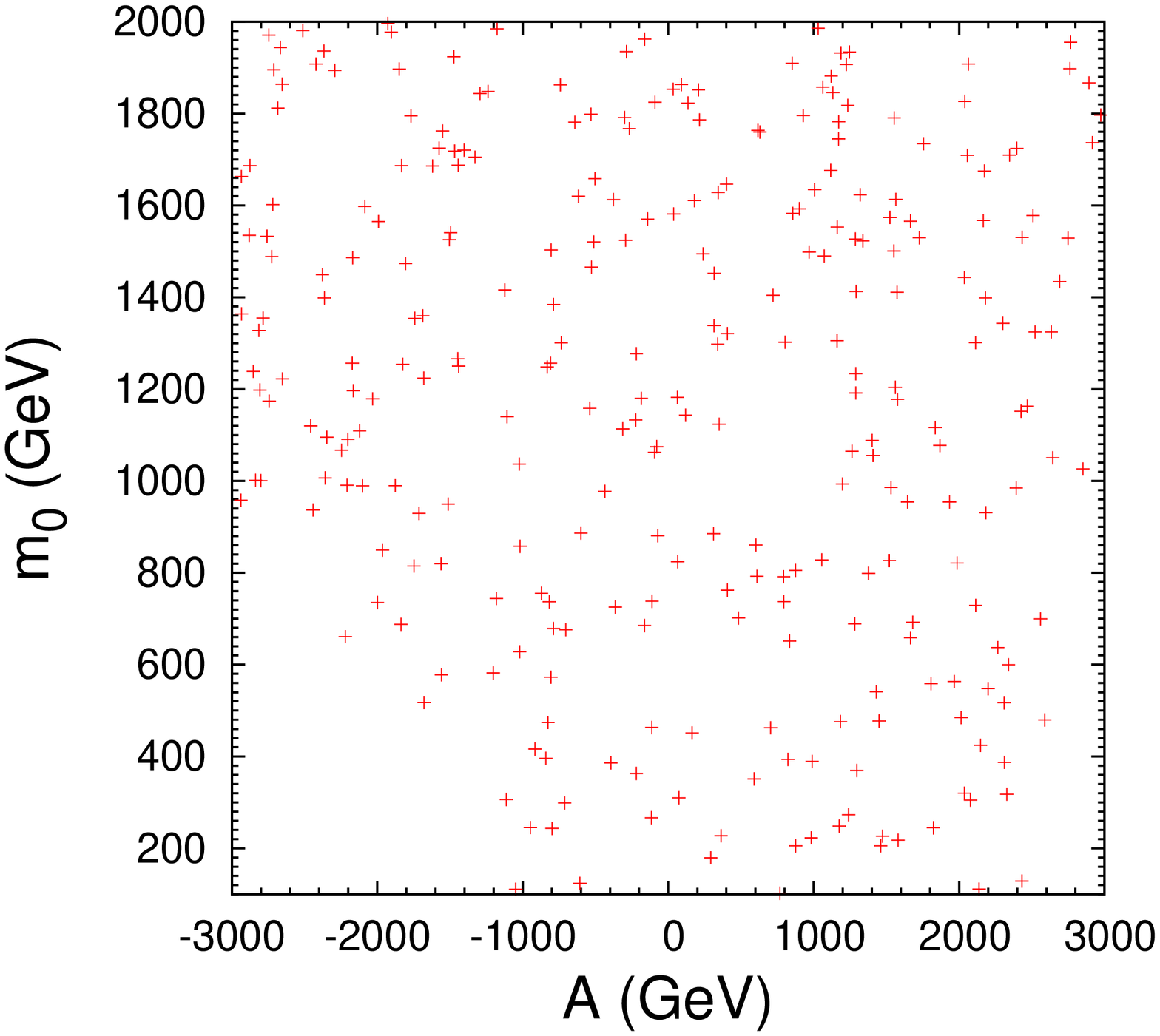}}
}
\caption{\small\sf RPV $(A-m_0)$ plane for $\mu>0$ for $\br=(5.8,\,0.50,\,0.10) \times 10^{-8}$.}
\label{fig:rpv2}
\end{figure}
\begin{figure}
\centerline{
\subfloat[$5.8\times10^{-8}$]{\label{fig:asmssv0}\includegraphics[angle=-90, width=0.35\textwidth]{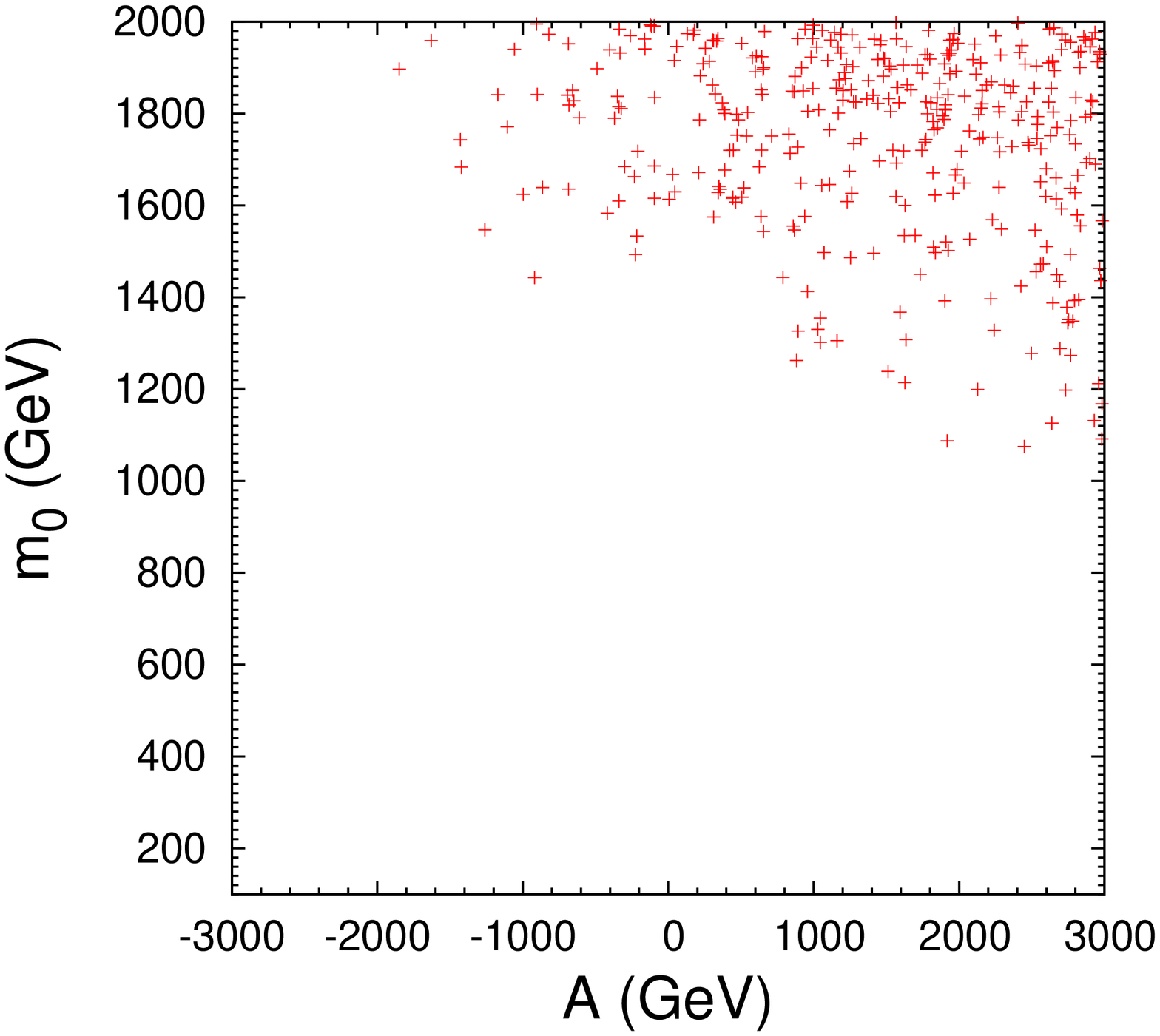}}  
\subfloat[$1.0\times10^{-8}$]{\label{fig:asmssv1}\includegraphics[angle=-90, width=0.35\textwidth]{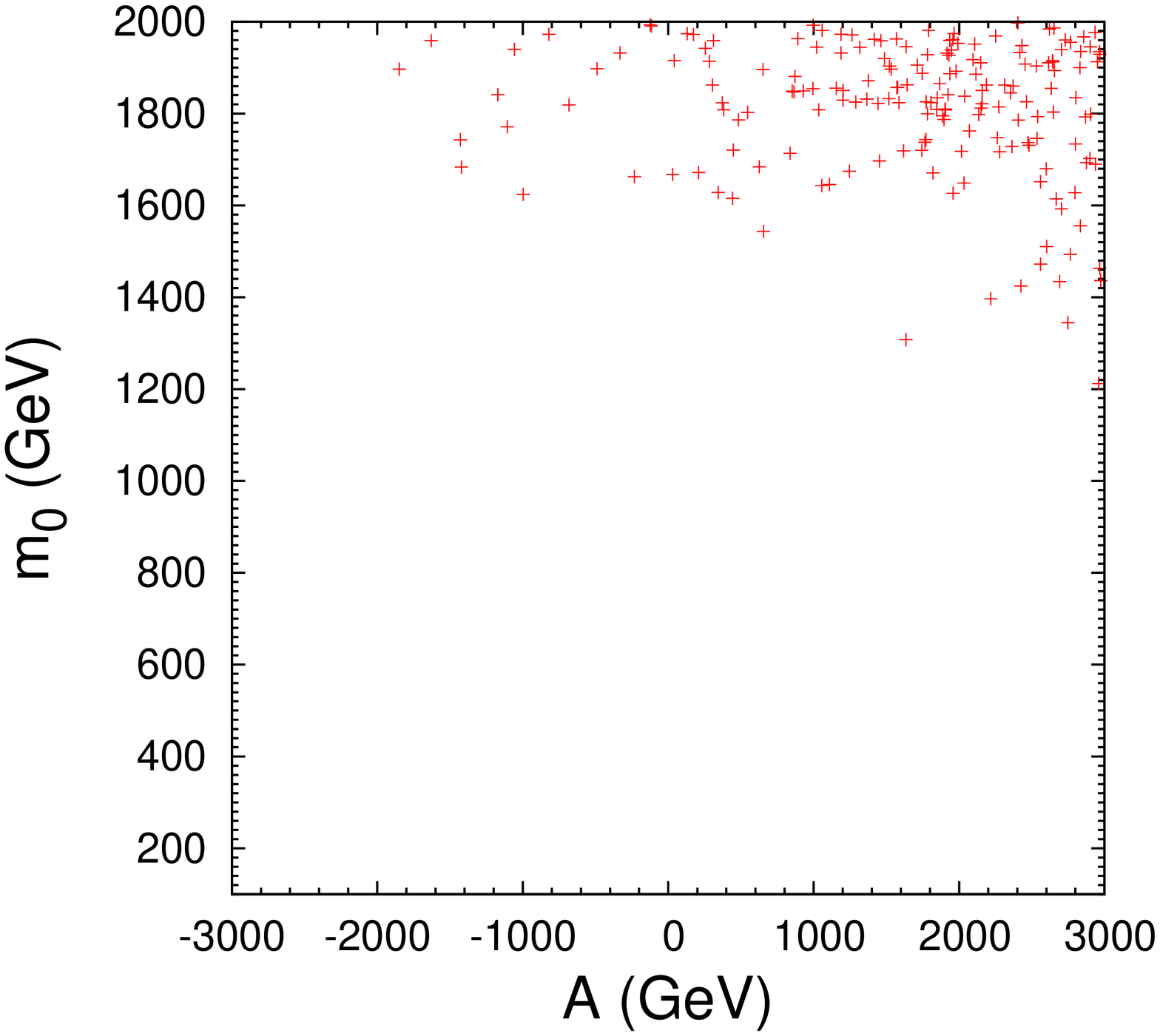}}
\subfloat[$5.0\times10^{-9}$]{\label{fig:asmssv2}\includegraphics[angle=-90, width=0.35\textwidth]{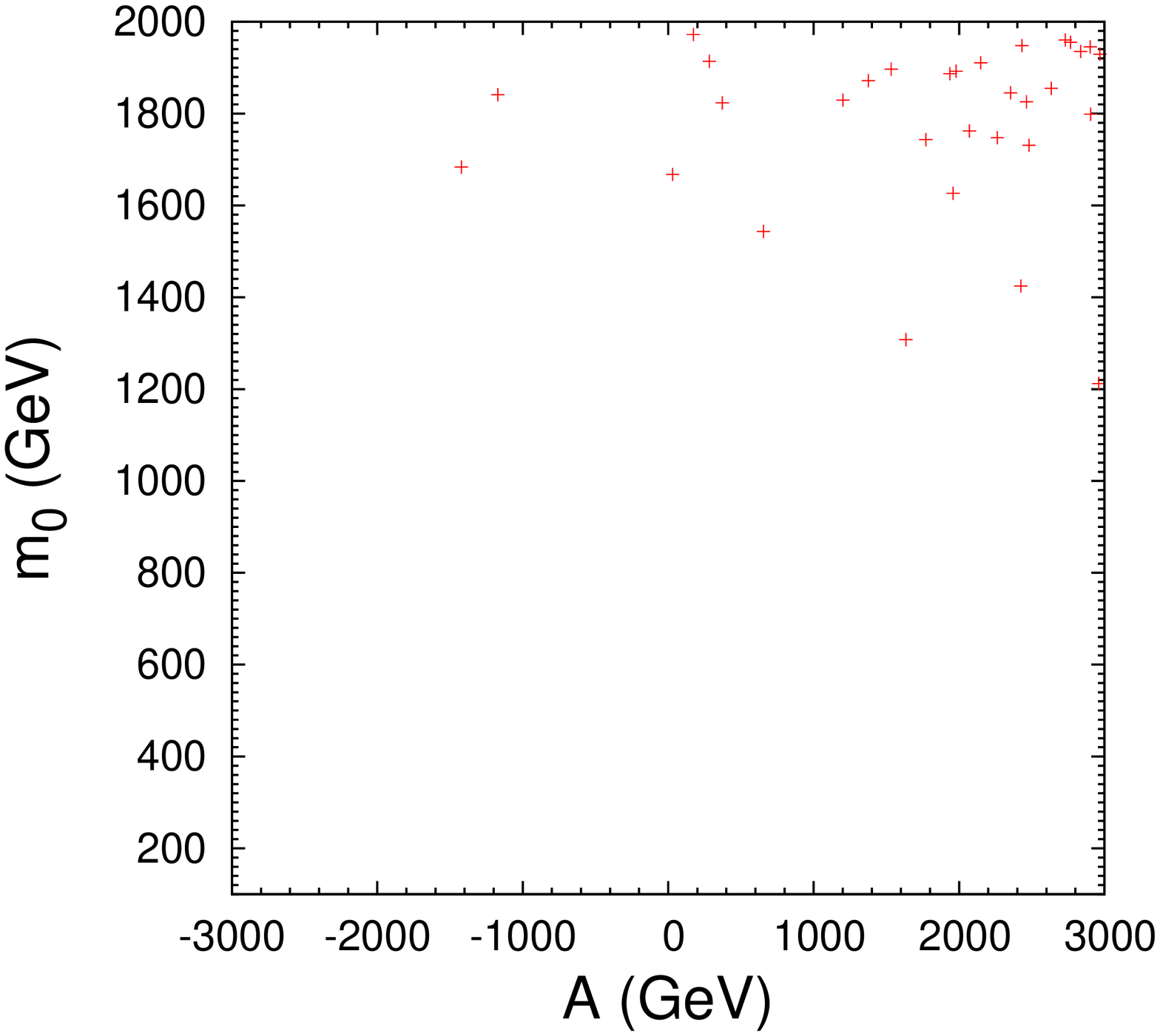}}
}
\caption{\small\sf RPV $(A-m_0)$ plane for $\mu<0$ for $\br=(5.8,\,1.0,\,0.50) \times 10^{-8}$.}
\label{fig:rpvs2}
\end{figure}
The constraints on the allowed RPV $(A-m_{0})$ plane for $\mu>0$ and $\mu<0$ 
are shown in Figure \ref{fig:rpv2} and \ref{fig:rpvs2} respectively. 
For $\mu>0$ , the present upper bound on the branching ratio of $\bs$ 
fails to put any useful constraint on the RPV $(A-m_{0})$ plane whereas 
for $\mu<0$  all parameter space for $m_0 \lesssim 1000\, \gev$ is ruled out. 
If the upper bound on $\br$  is improved to a value close to its SM prediction, the constraints on $(A-m_{0})$ plane 
for $\mu>0$ remains almost the same whereas for  $\mu<0$,  almost all parameter space is ruled out.

\begin{figure}
\centerline{
\subfloat[$5.8\times10^{-8}$]{\label{fig:asmfv0}\includegraphics[angle=-90, width=0.35\textwidth]{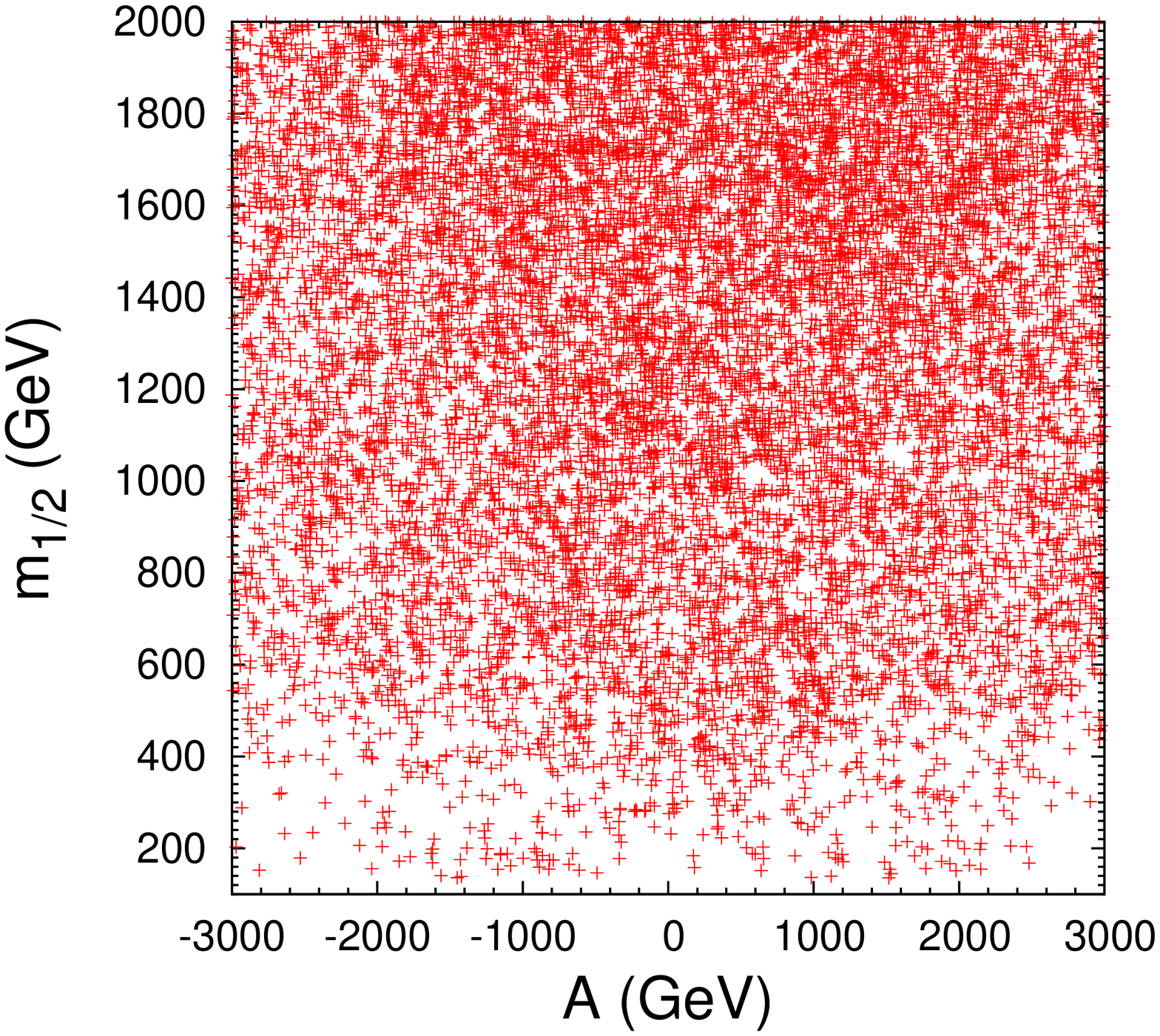}}  
\subfloat[$5.0\times10^{-9}$]{\label{fig:asmfv1}\includegraphics[angle=-90, width=0.35\textwidth]{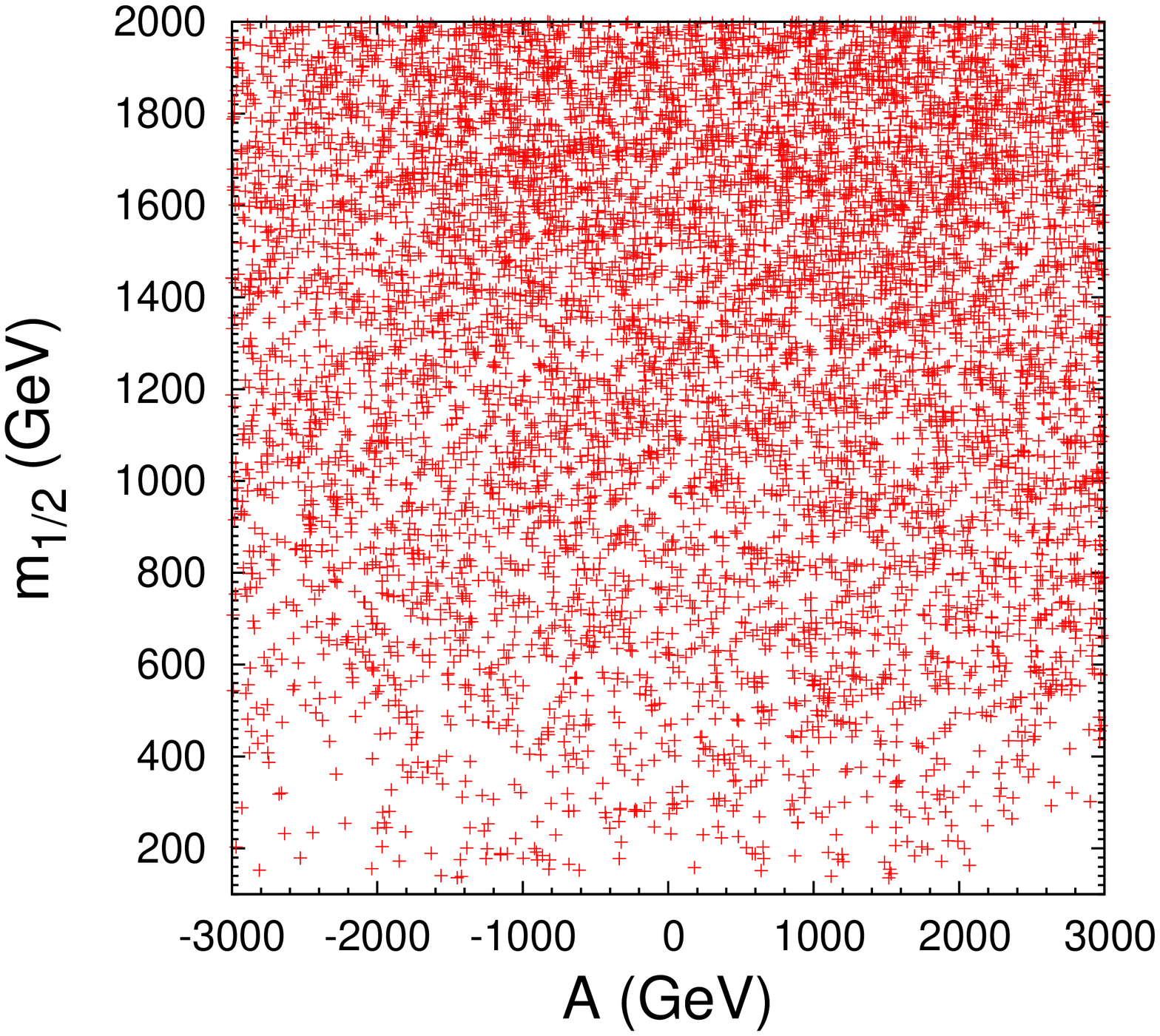}}
\subfloat[$1.0\times10^{-9}$]{\label{fig:asmfv2}\includegraphics[angle=-90, width=0.35\textwidth]{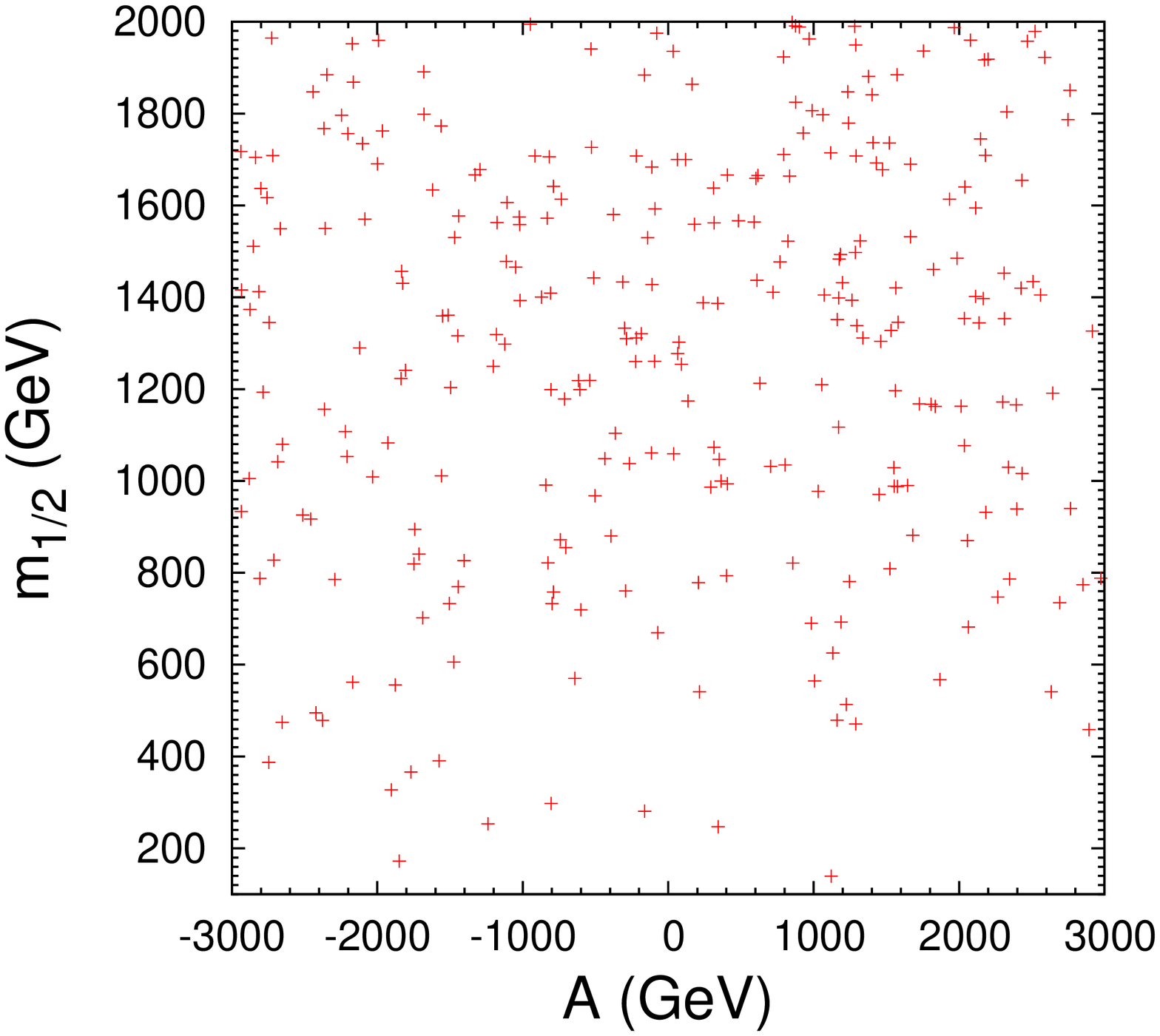}}
}
\caption{\small\sf RPV $(A-m_{1/2})$ plane for $\mu>0$ for $\br=(5.8,\,0.50,\,0.10) \times 10^{-8}$.}
\label{fig:rpv3}
\end{figure}
\begin{figure}
\centerline{
\subfloat[$5.8\times10^{-8}$]{\label{fig:asmfsv0}\includegraphics[angle=-90, width=0.35\textwidth]{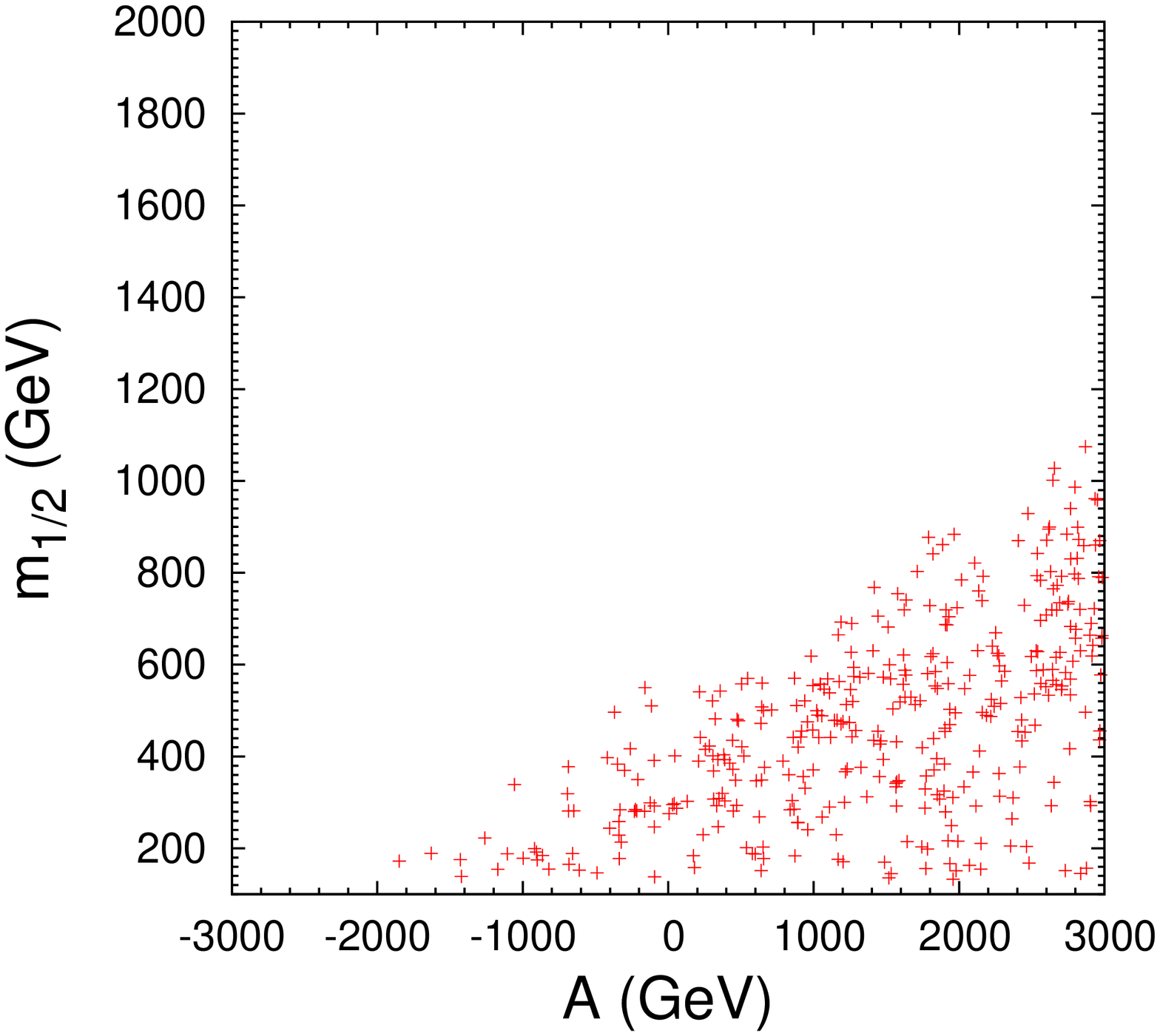}}  
\subfloat[$1.0\times10^{-8}$]{\label{fig:asmfsv1}\includegraphics[angle=-90, width=0.35\textwidth]{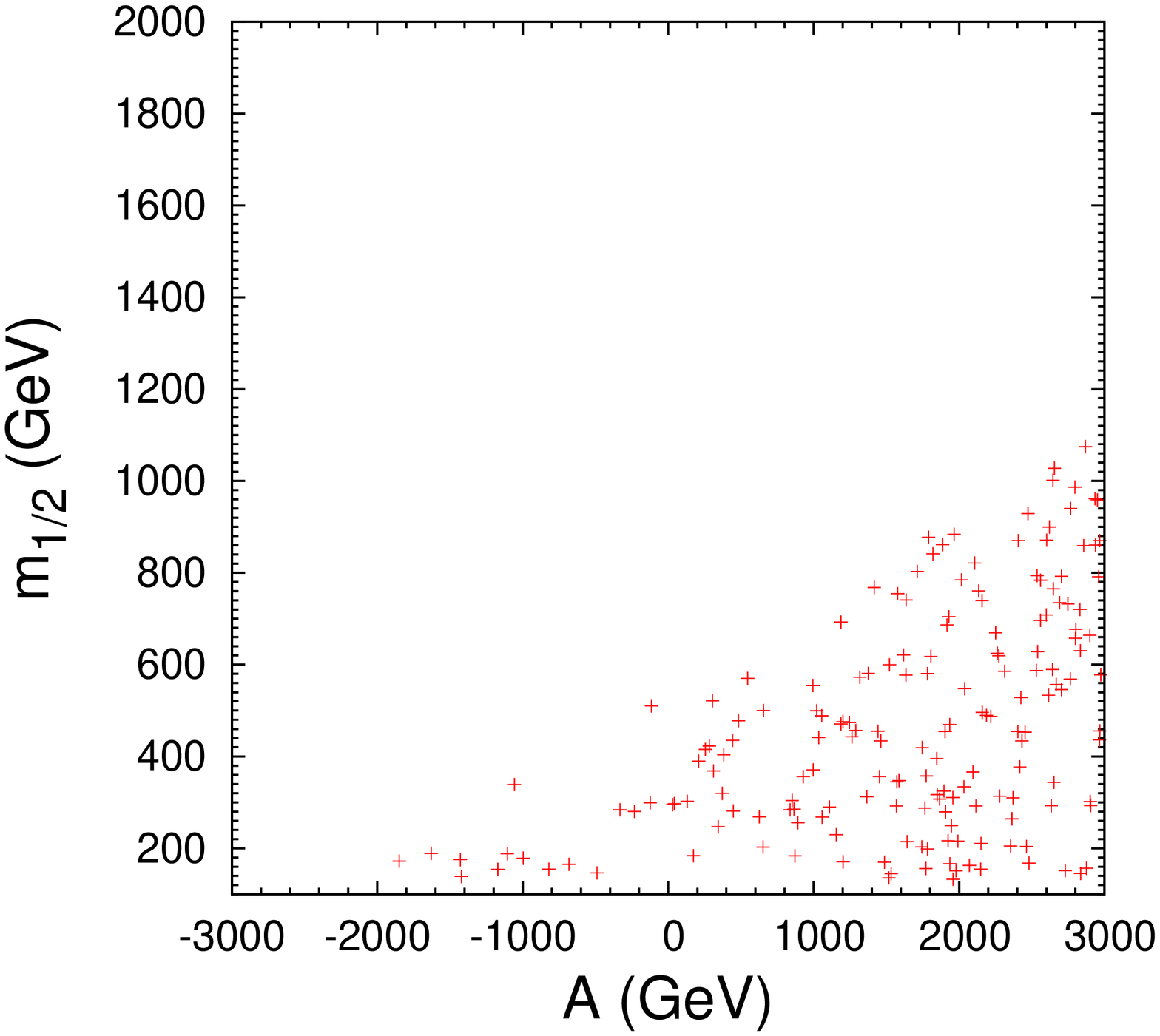}}
\subfloat[$5.0\times10^{-9}$]{\label{fig:asmfsv2}\includegraphics[angle=-90, width=0.35\textwidth]{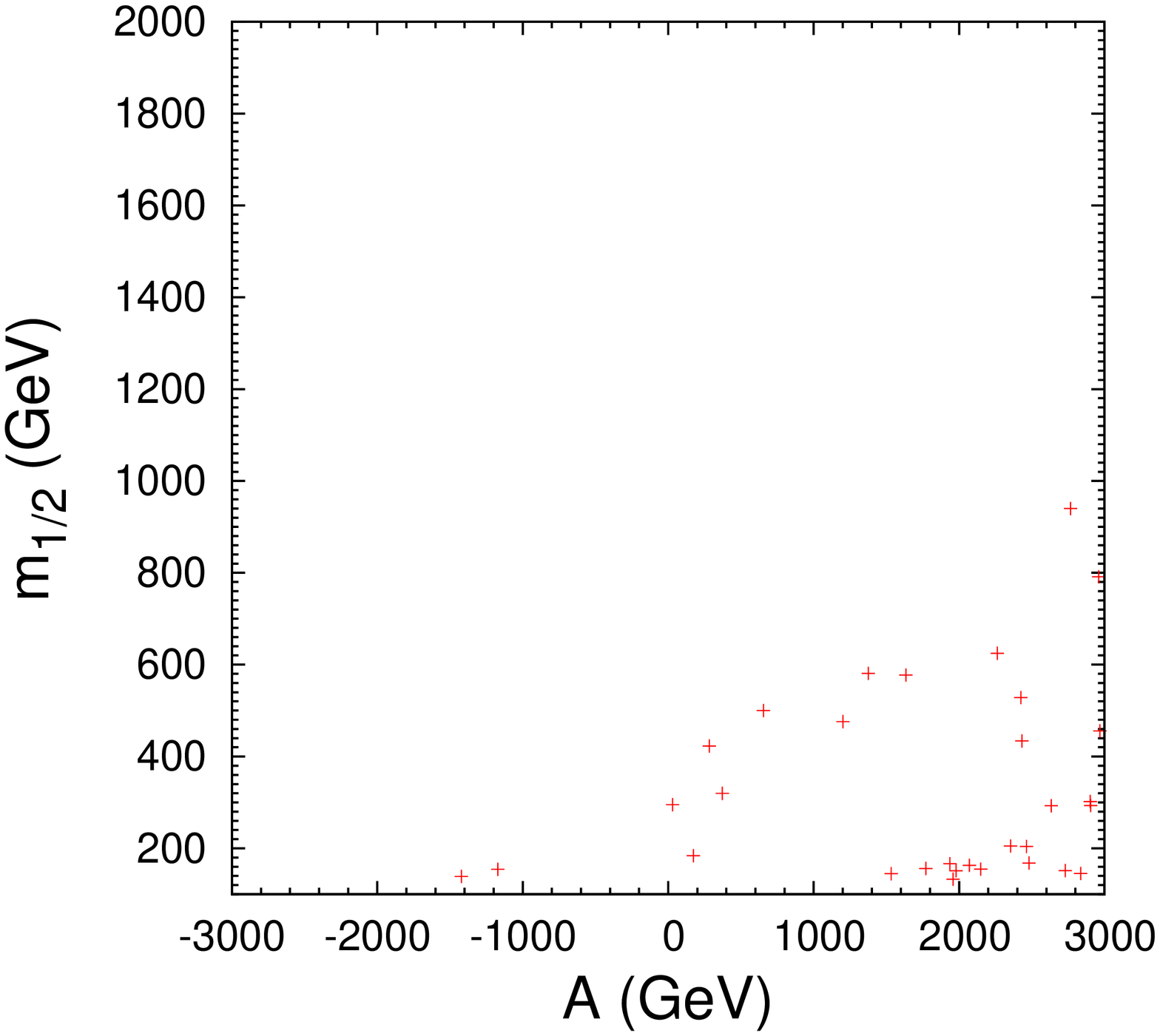}}
}
\caption{\small\sf RPV $(A-m_{1/2})$ plane for $\mu<0$ for $\br=(5.8,\,1.0,\,0.50) \times 10^{-8}$.}
\label{fig:rpvs3}
\end{figure}
Figure \ref{fig:rpv3} shows the allowed RPV $(A-m_{1/2})$ plane for 
$\mu>0$ corresponding to several possible values of the branching ratio of $\bs$. It is obvious 
from the Figure \ref{fig:asmfv0} that the present upper bound on the branching ratio of $\bs$ fails to put any useful constraint on the $(A-m_{1/2})$ plane. The situation remains almost the same even if the upper bound is improved by an order of magnitude. However a large parameter space is ruled out if the upper bound on  $\bs$ is as low as $1.0 \times 10^{-9}$. For $\mu<0$, the allowed RPV $(A-m_{1/2})$ plane is shown in Figure \ref{fig:rpvs3}. It can been seen that the constraints on 
$(A-m_{1/2})$ plane are more severe as compared to the case when $\mu>0$. For the present upper bound on $\br$, there is no allowed region for $m_{1/2} \gtrsim 1000\, \gev$. For $\br$ close to the SM prediction, almost all $(A-m_{1/2})$ parameter space is ruled out.

\begin{figure}
\centerline{
\subfloat[$5.8\times10^{-8}$]{\label{fig:tbmhv0}\includegraphics[angle=-90, width=0.35\textwidth]{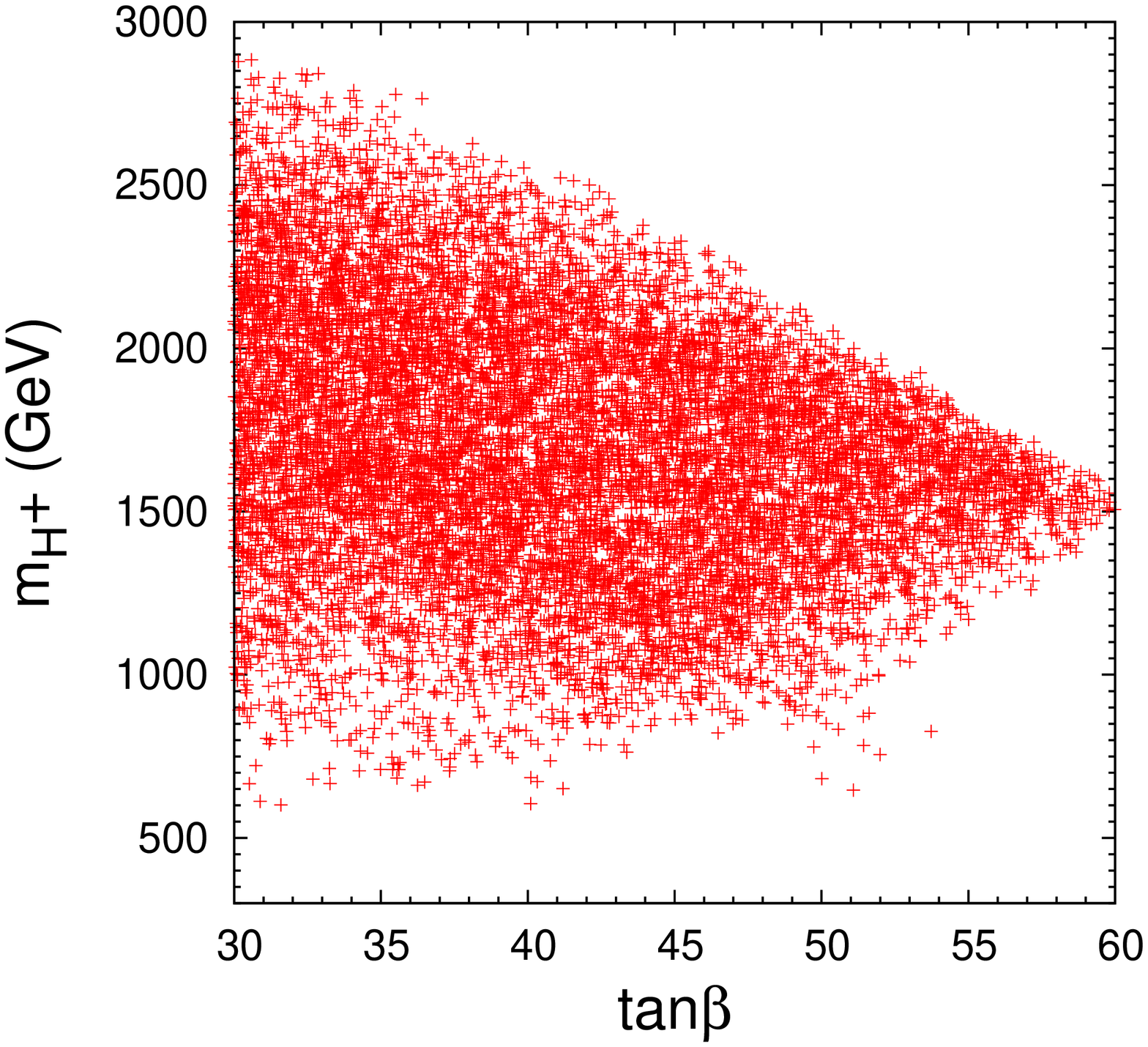}}  
\subfloat[$5.0\times10^{-9}$]{\label{fig:tbmhv1}\includegraphics[angle=-90, width=0.35\textwidth]{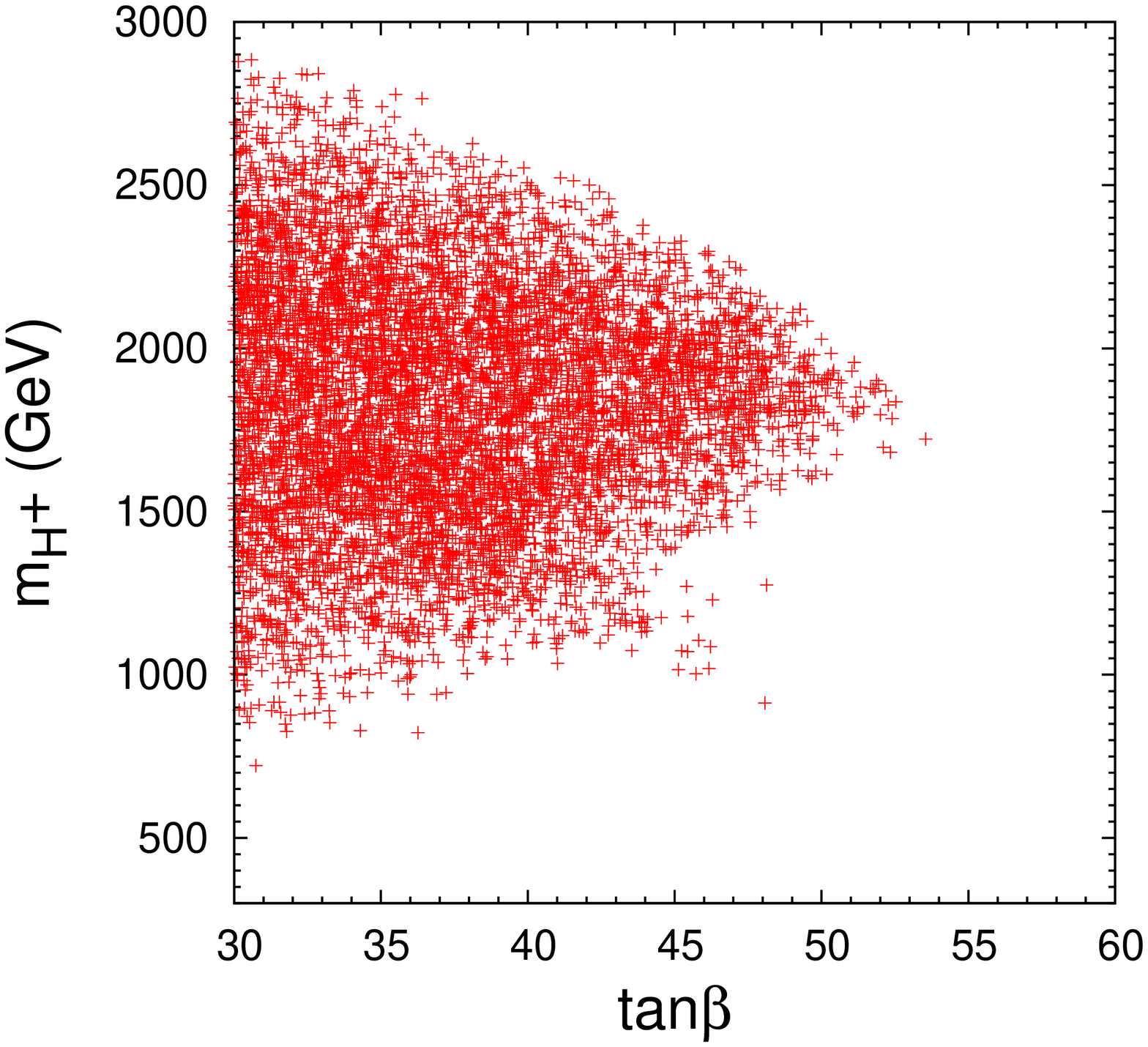}}
\subfloat[$1.0\times10^{-9}$]{\label{fig:tbmhv2}\includegraphics[angle=-90, width=0.35\textwidth]{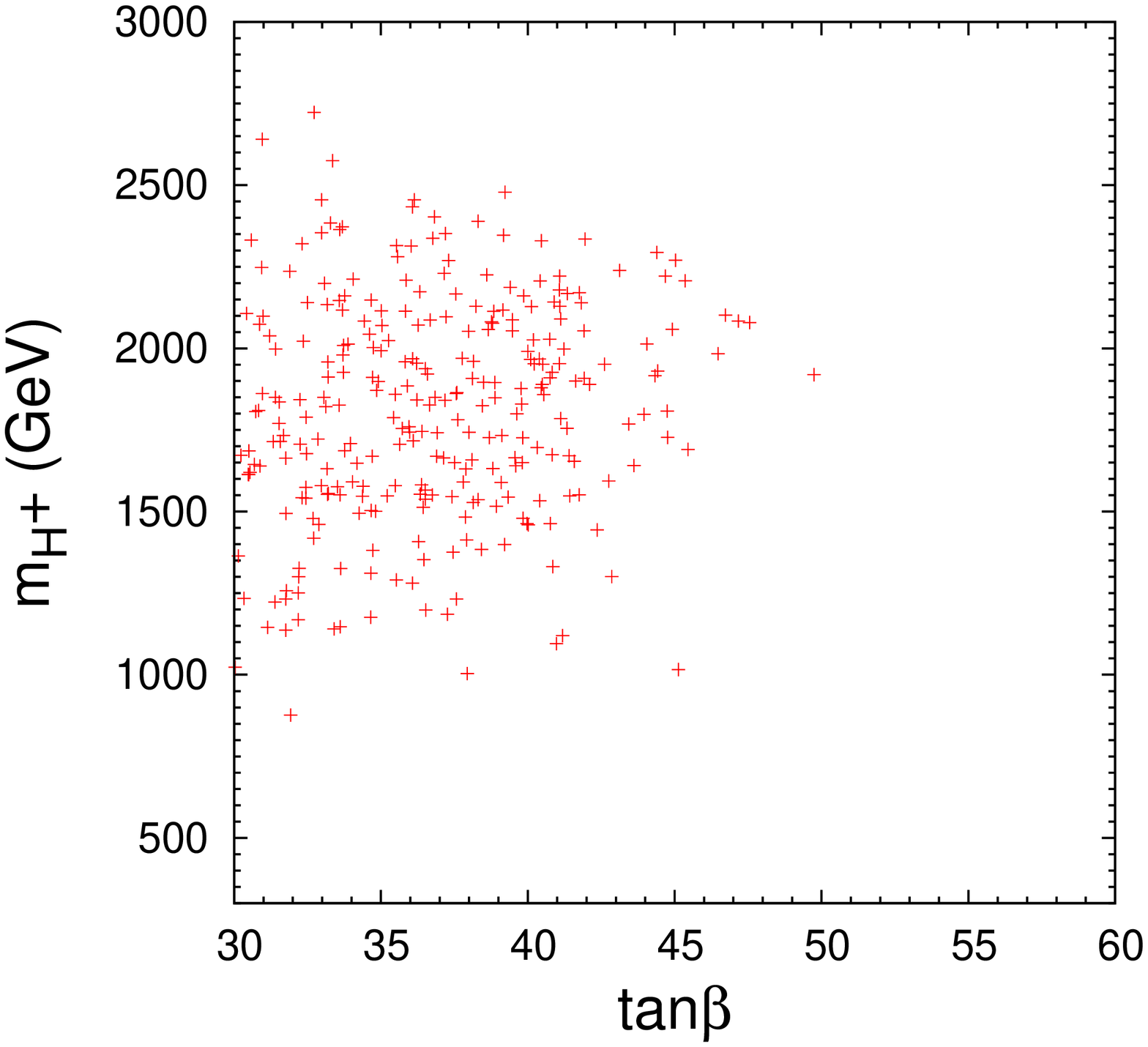}}
}
\caption{\small\sf RPV $(m_{H^\pm}-\tan\beta)$ plane for $\mu>0$ for $\br=(5.8,\,0.50,\,0.10) \times 10^{-8}$.}
\label{fig:rpv4}
\end{figure}
\begin{figure}
\centerline{
\subfloat[$5.8\times10^{-8}$]{\label{fig:tbmhsv0}\includegraphics[angle=-90, width=0.35\textwidth]{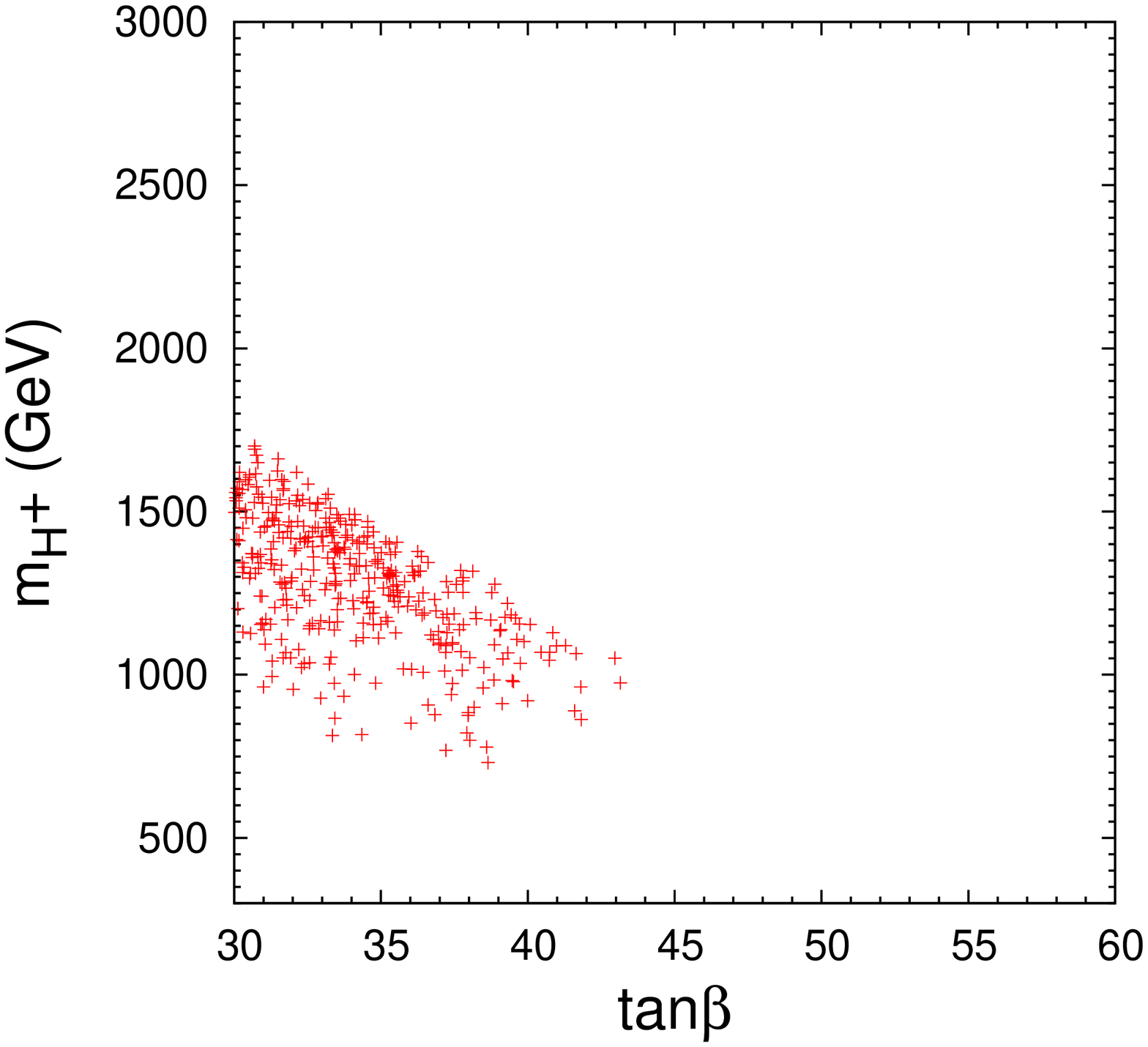}}  
\subfloat[$1.0\times10^{-8}$]{\label{fig:tbmhsv1}\includegraphics[angle=-90, width=0.35\textwidth]{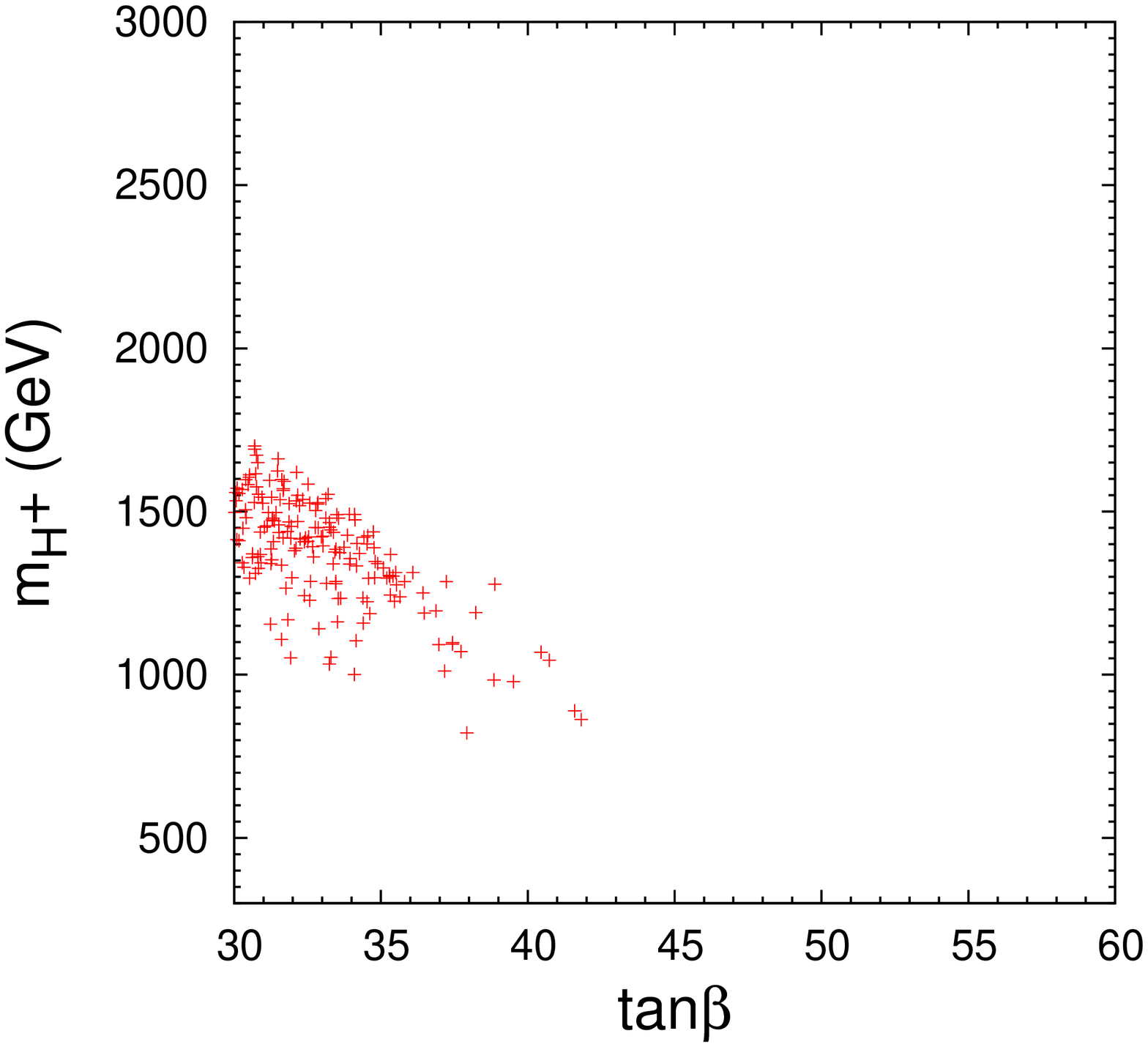}}
\subfloat[$5.0\times10^{-9}$]{\label{fig:tbmhsv2}\includegraphics[angle=-90, width=0.35\textwidth]{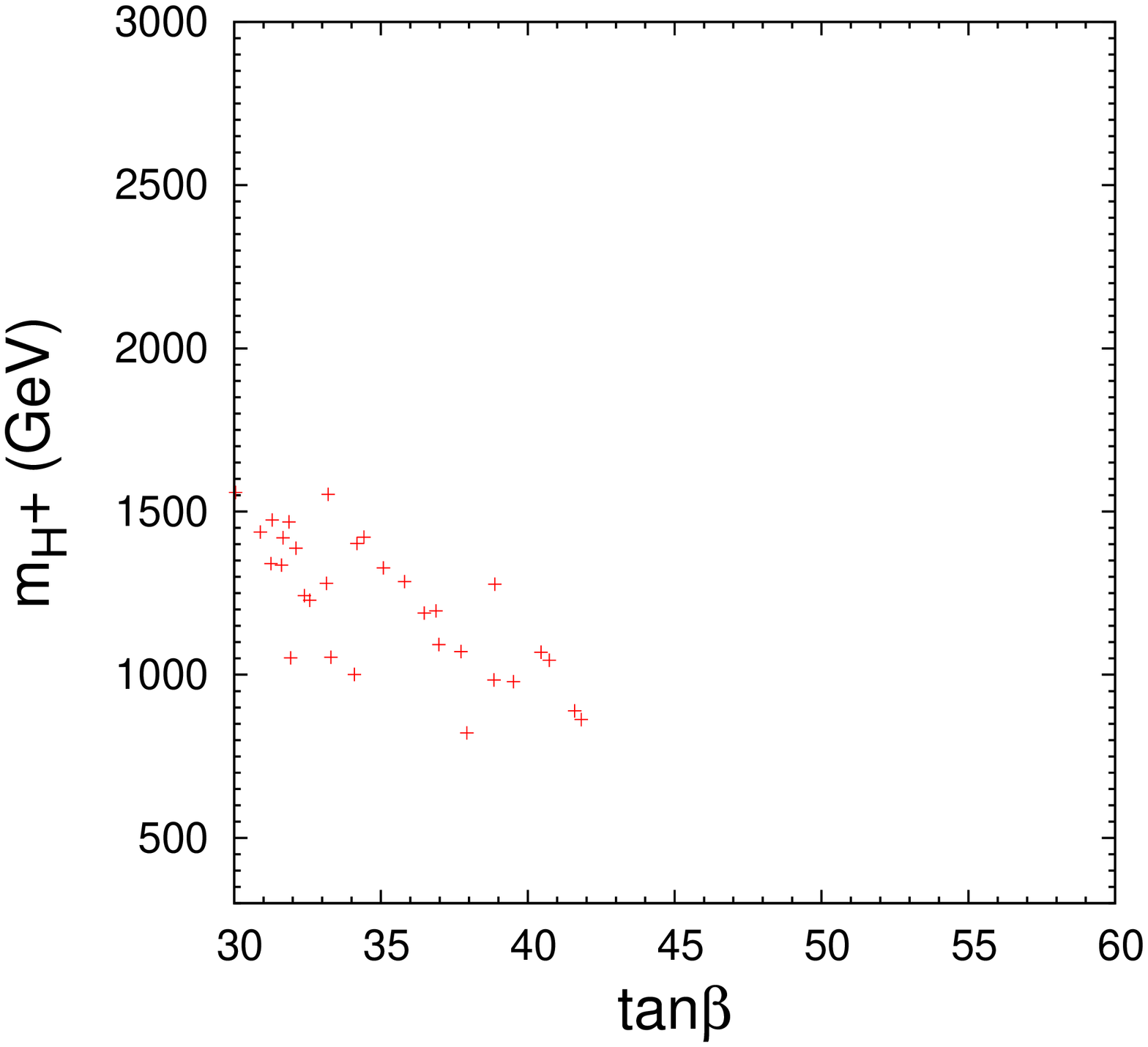}}
}
\caption{\small\sf RPV $(m_{H^\pm}-\tan\beta)$ plane for $\mu<0$ for $\br=(5.8,\,1.0,\,0.50) \times 10^{-8}$.}
\label{fig:rpvs4}
\end{figure}
Figure \ref{fig:rpv4}  shows the allowed $(m_{H^\pm}-\tan\beta)$ parameter space in the RPV mSugra for $\mu>0$. 
We see that the present upper bound on the branching ratio of $\bs$ puts a useful constraint on the $(m_{H^\pm}-\tan\beta)$ parameter space. 
The constraints of RPV Sugra parameter space becomes slightly weaker in comparison to that of R-parity conserving case is due to additional (RPV) parameters. The constraints are severe if the upper bound on $\br$ is  close to the LHCb sensitivity. In this case all  $(m_{H^\pm}-\tan\beta)$ parameter space is ruled out for $\tan\beta \gtrsim 50$. If $\mu<0$, then it can seen from the Figure \ref{fig:rpvs4} that for the present upper bound on $\br$, whole $(m_{H^\pm}-\tan\beta)$ parameter space for $\tan\beta \gtrsim 45$ is ruled out. For $\br$ close the SM prediction, almost all $(m_{H^\pm}-\tan\beta)$ parameter space is ruled out.

We noted that $\mu<0$ is strongly constraint as compared to that of $\mu>0$. This is due to the fact that contribution due to threshold correction depends on $sign(\mu)$ \cite{Allanach:2009ne}.  For $\mu<0$ the Higgs mass gets negative contribution due to the threshold corrections, thus making it inconsistent with the existing bound on the charged Higgs Mass in most of the $\mu<0$ region. We therefore observe relatively 
much smaller allowed parameter space for $\mu<0$ (see the Figures \ref{fig:rpv4} and \ref{fig:rpvs4}).

\section{Conclusions}
\label{concl}

In this paper we study the decay $\bs$ in context of the RPV mSugra in 
the high $\tan\beta$ regime. In a mininal flavour violating scenario, we 
consider contribution from the two Higgs doublet and the RPC terms along 
with the RPV terms. The results may be summarized as follows:

We show that even in the case of large $\tan\beta$, the lowest possible value of the branching 
ratio of $\bs$ in the RPV mSugra can go several 
orders of magnitude below the present LHCb sensitivity, and hence $\bs$ may be invisible to the LHC.

We find that the present upper bound on $\br$ puts strong constraint on 
the $\mu<0$ mSugra parameter space. Almost whole RPV mSugra parameter space, except for a 
few points below $\tan\beta = 50$ becomes disfavored. Once the constraint on $\bs$ are 
brought down to a value close to the projected LHCb 
sensitivity, there is hardly any allowed region. For $\mu>0$, constraints are 
relatively weaker.

\subsection*{Acknowledgements}

We thank A. Dighe and S. Uma Sankar for helpful discussions and S. 
Raichaudhuri for technical help. SKG acknowledges the Tata Institute of 
Fundamental Research, Mumbai for their hospitality during initial stage 
of the work. This work was partially supported by funding available from 
the US Department of Energy (DOE) under the contract number 
DE-FG02-01ER41155.

%
%

%
\end{document}